\documentclass[floatfix,preprint]{revtex4}
\usepackage{graphicx}
\usepackage{dcolumn}
\usepackage{color}
\usepackage{array}

\begin{document}

\title{Reflection and implantation of low energy helium with tungsten surfaces}
\author{Valery Borovikov}
\email{valery@lanl.gov} 
\affiliation{Theoretical Division, Los Alamos National Laboratory, Los Alamos, NM, USA}
\author{Arthur F. Voter}
\email{afv@lanl.gov} 
\affiliation{Theoretical Division, Los Alamos National Laboratory, Los Alamos, NM, USA}
\author{Xian-Zhu Tang}
\email{xtang@lanl.gov}
\affiliation{Theoretical Division, Los Alamos National Laboratory, Los Alamos, NM, USA}

\date{\today}

\begin{abstract}
  Reflection and implantation of low energy helium (He) ions by
  tungsten (W) substrate are studied using molecular dynamics (MD)
  simulations.  Motivated by the ITER divertor design, our study
  considers a range of W substrate temperatures (300~K, 1000~K,
  1500~K), a range of He atom incidence energies ($\le$100~eV) and a
  range of angles of incidence ($0^{\circ}$-$75^{\circ}$) with respect
  to substrate normal. The MD simulations quantify the reflection and
  implantation function, the integrated moments such as the
  particle/energy reflection coefficients and average implantation
  depths.  Distributions of implantation depths, reflected energy,
  polar and azimuthal angles of reflection are obtained, as functions
  of simulation parameters, such as W substrate temperature, polar
  angle of incidence, the energy of incident He, and the type of W
  substrate surface.  Comparison between the MD simulation results,
  the results obtained using SRIM simulation package, and the existing
  experimental and theoretical results is provided.

\end{abstract}

\maketitle

\section{Introduction}

Plasma surface interaction (PSI) at the first wall and divertor of a
fusion reactor not only poses a materials challenge in terms of
radiation damage due to the extreme plasma and neutron irradiation
flux, but also introduces a controlling factor on the boundary plasma
conditions through recycling and impurity production. The latter
effect has a direct impact on fusion power output since edge plasma is
known to affect the performance of fusion-producing core plasmas.
Presently both solid wall and liquid wall designs are being actively studied.
The design of ITER chooses a combination of beryllium (Be) first wall
and tungsten (W) divertor. A number of current tokamak upgrades and
proposed future machines have been exploring a configuration of
all-tungsten first wall and divertors.  The primary benefits are the
high melting temperature, the low sputtering yield, and the excellent
thermal conductivity of refractory metals in general and tungsten in
particular~\cite{Philipps}.

The plasma ion irradiation flux to the tungsten surface in a fusion
reactor is made of deuteron (D) and triton (T), which are unburned
fusion fuel, and helium (He), which is the fusion product.  Unlike D
and T, helium is chemically inert, but can bring severe damage to the
tungsten surface by clustering in the form of subsurface bubbles,
which can burst and induce complicated surface morphology known as
fuzz.  Also unlike D and T, helium can not aggregate on the tungsten
surface as a deposited He layer. The incoming helium is either
implanted below the W surface or reflected upon impact.  The implanted
He can migrate to the tungsten surface on a diffusive time scale.  The
desorption of He from a tungsten surface is energetically favorable so
a mostly clean tungsten surface is maintained in a He plasma.  In a
working fusion reactor, the recycling of He will be complicated by the
presence of triton and deuteron at the tungsten surface.  Here we will
focus on the pure He plasma situation. This simplification is helpful
in establishing a better understanding of the fundamental process of He
ion interaction with a tungsten surface. It is also practically
important because, due to heating power constraint for accessing high (H)
confinement mode, there is a proposed pure He plasma phase in the ITER
program start-up.

Our emphasis here is on low energy helium ions, in the range of 1-100
electron volts (eV) at the time of impact with the tungsten surface.
This is to be consistent with the design goal/choice that takes
advantage of the negligibly small sputtering yield of light ions on
tungsten surface at low energy.  For an impacting He ion with energy
$E_i,$ incident polar angle $\theta_i$ and azimuthal angle
$\varphi_i,$ the probability of it being reflected by the tungsten
surface with an outgoing energy $E_r$ into a differential solid angle
$d\Omega \equiv \sin\theta_r d\theta_r d\varphi_r$ about $\theta_r$
and $\varphi_r,$ is ${\cal T}(E_r, \theta_r, \varphi_r | E_i, \theta_i,
\varphi_i) d\Omega.$ Evidently ${\cal T}(E_r, \theta_r, \varphi_r |
E_i, \theta_i, \varphi_i; \Sigma, T)$ is also a function of the
tungsten surface type $\Sigma$ and its temperature $T.$ The
total probability of the He ion being reflected upon impact is
\begin{equation}
\label{eq:reflection-coeff}
R(E_i, \theta_i, \varphi_i; \Sigma, T) =
\int_0^\infty dE_r \int_0^{\pi/2} \sin\theta_r d\theta_r \int_0^{2\pi} d\varphi_r
{\cal T},
\end{equation}
and the average energy of the reflected He is given by
\begin{equation}
\left<E_r\right> 
=
\int_0^\infty dE_r \int_0^{\pi/2} \sin\theta_r d\theta_r \int_0^{2\pi} d\varphi_r
E_r {\cal T}.
\end{equation}
The so-called energy reflection coefficient for mono-energetic He ion striking the
tungsten at a particular orientation is given by
\begin{equation}
\label{eq:energy-reflect-coeff}
R_E(E_i,\theta_i,\varphi_i; \Sigma, T) 
\equiv \frac{\left<E_r\right>}{E_i}.
\end{equation}

The reflection function ${\cal T}$ provides the energy and angular
distribution of the reflected He atoms
$F_r(E_r,\theta_r,\varphi_r)$ for any given impacting He ion
distribution $F_i(E_i,\theta_i,\varphi_i),$
\begin{widetext}
\begin{equation}
  F_r(E_r,\theta_r,\varphi_r)
  = \int_0^\infty dE_i \int_0^{\pi/2} \sin\theta_i d\theta_i \int_0^{2\pi} d\varphi_i
  {\cal T}(E_r, \theta_r, \varphi_r | E_i, \theta_i, \varphi_i; \Sigma, T) F_i(E_i,\theta_i,\varphi_i).
\end{equation}
\end{widetext}
$F_r$ provides the wall feedback flux to the boundary plasma, which
is described by a set of combined plasma/neutral evolution equations.
In other words, $F_r$ sets the boundary condition for a boundary
plasma model that can inform us on the energy and angular distribution
of the impacting He ions.  The importance of this recycling process is
that it provides a crucial link in the feedback loop of PSI that governs the
boundary plasma condition in a fusion reactor.

Particle conservation implies that $1 - R(E_i,\theta_i,\varphi_i;
\Sigma, T)$ is the probability of the impacting ion being implanted
into the tungsten.  For the implanted He ions, the final resting
position gives the so-called range distribution function ${\cal
  S}(l,\theta_s,\varphi_s | E_i, \theta_i, \varphi_i; \Sigma, T)$ with
$l$ the distance from the surface,
$\theta_s$ the polar angle and $\varphi_s$ the azimuthal angle, all
relative to the point of initial impact at the tungsten surface.  By
definition,
\begin{widetext}
\begin{equation}
  \int_0^\infty l^2dl \int_0^{\pi/2} \sin\theta_s d\theta_s \int_0^{2\pi} d\varphi_s
  {\cal S}(l,\theta_s,\varphi_s | E_i, \theta_i, \varphi_i; \Sigma, T) 
  = 1 - R(E_i,\theta_i,\varphi_i;\Sigma,T).
\end{equation}
\end{widetext}
The average projected range or implantation depth normal to the surface 
for an impacting He ion of
$(E_i,\theta_i,\varphi_i)$ is
\begin{equation}
\label{eq:average-range}
L(E_i,\theta_i,\varphi) =
\int_0^\infty l^2 dl\int_0^{\pi/2}\sin\theta_s d\theta_s \int_0^{2\pi}d\varphi_s x {\cal S},
\end{equation}
where
\begin{equation}
x \equiv l \cos\theta_s
\end{equation}
the projected range.
For an impacting He distribution of $F_i(E_i,\theta_i,\varphi_i),$ the
total range distribution $S(l,\theta_s,\varphi_s)$ is
\begin{widetext}
\begin{equation}
  S(l,\theta_s,\varphi_s)
  = \int_0^\infty dE_i \int_0^{\pi/2} \sin\theta_i d\theta_i \int_0^{2\pi} d\varphi_i
  {\cal S}(l, \theta_s, \varphi_s | E_i, \theta_i, \varphi_i; \Sigma, T) 
  F_i(E_i,\theta_i,\varphi_i).
\end{equation}
\end{widetext}

The range distribution $S(l,\theta_s,\varphi_s),$ or equivalently,
$S(x,\theta_s,\varphi_s),$
provides the source information of implanted He
atoms for understanding their eventual effect on the bulk and surface
properties of the tungsten divertor and first wall.  This is the
materials side of plasma-surface interaction, in contrast to
$F_r(E_r,\theta_r,\varphi_r)$ which provides the influence of
plasma-surface interaction on the plasmas.  Specifically for tungsten,
implantation of He can cause He-bubble formation, blistering and
formation of W nanostructure (``fuzz") on the
surface \hspace{1mm}\cite{Tokitani, Yoshida, Kajita}. This can lead to
rapid erosion and significant degradation of the mechanical properties
and heat load resistance of the material.  Interaction of He with W
surfaces has been investigated experimentally and theoretically for
the last four decades~\cite{vanGorkum, Amano, Robinson, Tabata,
  Eckstein, Thomas}.  There were also a number of simulation
studies. In particular, Henriksson {\it et al}.\hspace{1mm}(see
Refs.\hspace{1mm}\onlinecite{Henriksson1, Henriksson2}) studied He
bubble formation and initial stages of blistering in He implanted W.
Li {\it et al}.\hspace{1mm}(see Ref.\hspace{1mm}\onlinecite{Li})
studied temperature effects on low energy He bombardment of the W
surface. However, only the W(100) surface was considered and only 
bombardment with incident direction normal to the surface was studied.

The primary objective of this paper is to quantify ${\cal T}$ and
${\cal S}$ for low energy He ion bombardment of W surfaces.  This
information is required for further studies of wall recycling on
boundary plasmas (${\cal T}$) and plasma irradiation on materials
properties (${\cal S}$).  Molecular dynamics is arguably the method of
choice to provide ${\cal T}$ via direct numerical simulations. This
will become obvious as one encounters the detailed multiple multi-body
collision processes as opposed to single binary collision in He
reflection by W surfaces.  For the low energy He ion considered here,
MD is also an excellent tool for quantifying the range distribution
for the He implantation since the contribution from electronic
stopping is small.  We have performed MD simulations of low energy He
bombardment of three W surfaces, using three W substrate temperatures,
a range of He incident energies and a range of angles of incidence
with respect to substrate normal.  We have also compared the results
of MD simulations with the results obtained using the SRIM simulation
package\hspace{1mm}\cite{SRIM}, as well as with the existing
experimental and theoretical results. For the purpose of completeness
in documenting simulation data of ${\cal T}$ and ${\cal S}$ as
boundary and initial conditions for boundary plasma modeling and studies of He
transport in tungsten, we provide an unusually large amount of tables and plots.

The rest of the paper is organized as follows. The set up and
the procedures of our MD simulations are described in
Sec.~\ref{sec:method}.  The simulation results are given in
Sec.~\ref{sec:results}.  Specifically, a summary of the integrated
quantities for reflection and implantation, namely the particle and
energy reflection coefficients, and the average range of implantation, are
shown in Sec.~\ref{subsec:results-summary} as a function of substrate
temperature, surface type, incidence energy and angle.  The
range distribution is taken up in Sec.~\ref{subsec:results-range}.
The energy and angular distributions of the reflected He are examined
in Sec.~\ref{subsec:results-energy} and
Sec.~\ref{subsec:results-angular}, respectively.
For comparison, we also perform SRIM calculation in Sec.~\ref{subsec:results-srim},
and MD simulations, using a second EAM potential, 
in Sec.~\ref{subsec:results-derlet}.
Our results are contrasted with existing ones in Sec.~\ref{sec:comparison}, before
concluding remarks in Sec.~\ref{sec:conclusion}.

\section{Molecular dynamics simulation procedure\label{sec:method}}

Molecular dynamics is used in our atomistic studies of He reflection
and implantation with a tungsten surface.  To describe the interatomic
interaction between W atoms we used the Ackland-Thetford embedded
atom method (EAM) potential for W~\cite{Ackland}, modified
by Juslin {\it et al}~\cite{Juslin}. To describe the
interaction between He and W atoms we used the W-He pairwise potential
recently developed by Juslin {\it et al}~\cite{Juslin}. We
considered three bcc W surfaces ((100), (110), (310)), three
simulation temperatures (300~K, 1000~K,
1500~K), a range of He deposition energies
(0.5~eV - 100~eV), and a range of deposition
angles with respect to substrate normal ($0^{\circ}$-$75^{\circ}$).
For better statistics, we simulated 1000 He impacts for each
set of simulation parameters. In those cases where the initial
deposition angle with respect to substrate normal was not $0^{o}$, the
azimuthal angle for each deposition was drawn randomly (ranging from
$0^{\circ}$ to $360^{\circ}$).  This corresponds to a distribution of
impacting He ions $f_i$ that is independent of $\varphi_i,$ {\it i.e.},
$f_i = f_i(E_i,\theta_i).$

In the case of a W(100) surface, the size of
the simulation system varied from 6750 atoms 
($\sim$\hspace{0.2mm}47.5\hspace{0.2mm}\AA
$\times$\hspace{0.2mm}47.5\hspace{0.2mm}\AA
$\times$\hspace{0.2mm}47.5\AA) to 54000 atoms
($\sim$\hspace{0.2mm}95\hspace{0.2mm}\AA
$\times$\hspace{0.2mm}95\hspace{0.2mm}\AA
$\times$\hspace{0.2mm}95\AA), depending on the initial energy of the He
ion. In cases where the channeling effect\hspace{1mm}\cite{Nastasi} was especially pronounced, we
used a simulation system elongated in the z direction (up to
$\sim$\hspace{0.2mm}220\hspace{0.2mm}\AA ).  For (110) and (310)
surfaces we used similar system sizes.  Periodic boundary conditions
were applied in the x and y directions. In the z direction there are
free-standing top and bottom surfaces.  With relatively large
system sizes we did not need to use a thermostat for temperature control
during the MD simulations of He impact with W surfaces, because the
changes in temperature of the substrate under the influence of He
impact were comparable to regular temperature fluctuations.

Prior to MD simulations of He impact with the W surface we equilibrated
the substrate at the desired temperature by applying a Langevin
thermostat to all atoms. Non-accumulative He-impact MD simulations
were carried out for each set of deposition parameters (such as 
initial He impact energy $E_i$ and deposition angle $\theta_i$).  
In the beginning of each simulation a He atom was placed
above the substrate (the initial distance to the substrate was greater
than the W-He potential cutoff distance and x and y positions were
chosen randomly each time) and assigned a velocity with direction and
magnitude corresponding to the initial deposition angle and initial
deposition energy.
For each given deposition energy and temperature, for all 1000 trajectories
we used the same initially prepared substrate block to save computational
time.  In a test on the 40~eV, $T=1000$~K, normal-deposition case on the (100) surface,
we  verified using 5000 trajectories that this procedure gave the same results as a
procedure in which the substrate block was integrated forward in time by 0.5\hspace{0.2mm}ps
before each deposition was initiated.  
A time step of 0.1\hspace{0.2mm}fs was used in all
simulations.  In those cases where the He atom was reflected, we
interrupted the MD simulation immediately and collected the data
(reflected energy and angles of reflection). In cases of implantation
of He atom into the W substrate, the total MD simulation time was
3\hspace{0.2mm}ps. By the end of 3\hspace{0.2mm}ps MD simulation the
kinetic energy of a He atom is observed to always fall below the
magnitude of He migration energy in bcc W ($\sim
0.2$~eV)~\cite{Juslin}.

\section{Simulation Results\label{sec:results}}

\subsection{Summary of results for W(100), W(110), W(310) surfaces\label{subsec:results-summary}}

A summary of the MD simulation results can be conveniently given in
terms of the integrated quantities of particle reflection coefficient
$R(E_i,\theta_i),$ energy reflection coefficient $R_E(E_i,\theta_i),$
and average projected range $L(E_i,\theta_i)$ for the implanted He atoms.

In Table~\ref{table1} we present the summary results on particle and
energy reflection coefficients for low energy He bombardment of the W(100)
surface, for a substrate temperature of  $T=1000$~K. A range of
initial energies of He atoms is considered. As can be seen from
Table~\ref{table1}, there is a strong dependence of reflection
coefficient $R,$ energy reflection coefficient $R_E$ and implantation
depth $L$ on the initial incident energy of He atoms. We recall that a
reflection coefficient is the ratio of reflected He atoms to the total
number of impacting He atoms, Eq.~(\ref{eq:reflection-coeff}).  The energy
reflection coefficient is defined as the ratio of the kinetic energy
of the reflected He atom to the incident kinetic energy of the He atom,
Eq.~(\ref{eq:energy-reflect-coeff}). 
Due to the high atomic number of tungsten, the reflection coefficients in both
particle number and energy are much higher than those of a carbon or beryllium wall.
Conventionally these reflection coefficients are anticipated to increase 
as the incident angle becomes larger, following a functional dependence of
$\cos^n\theta_i$ with the index $n$ negative. 
In our MD simulations, one can see from Table~\ref{table1} 
that the fraction of reflected He atoms do tend to increase (for
all incident energies), as the deposition angle with respect to
substrate normal ($\theta_i$)
becomes large, $\theta_i > 45$ degrees. For the largest deposition angle we have considered
($75^{\circ}$) all He atoms are reflected (100\hspace{0.2mm}\%), so
the reflection coefficient is equal to 1 for all incident energies in
this case. In cases of small deposition angle, the reflection
coefficient is equal to 1 only for the lowest incident energies (up to
10~eV). The energy reflection coefficient
has a similar behavior with respect to incident angle, and lower deposition energies
produce higher energy reflection coefficients. 
 What is different from the previous results~\cite{Tabata2} is the
  non-monotonic dependence of particle/energy reflection coefficients
  with the incidence angle, showing a minimum at $\theta_i \approx
  30-45^\circ.$ This is likely a geometrical effect of the tungsten
  lattice.  It is known that at low energy, the interaction between an
  impacting ion and lattice atoms is dominated by multiple collisional
  events, so the lattice structure of the top few layers from the line
  of sight of the impacting ion can play an important role in particle
  reflection.  This is consistent with the well-known channeling
  effect of the lattice.  Around the angle of $30-45^{\circ}$ with
  respect to the W(100) surface normal, such a geometrical channeling
  effect appears to reach a maximum and results in a shallow minimum
  for the reflection coefficient. As the incident angle becomes even
  larger, the higher effective lattice atom density along the line of
  sight will become the dominant effect and produce much larger
  reflection coefficients in both particle number and energy.  We note
  that this discussion uses a W(100) surface as an example.  For
  other surfaces the local minima in particle/energy reflection
  coefficients may correspond to different deposition angles.

The substrate temperature (see Tables ~\ref{table2}, \ref{table3})
does not affect the results significantly, except for the implantation
depth of He atoms at lower (300~K) temperature. For
substrate T=300~K  the average implantation depth increases
substantially in the case of normal deposition of He atoms due to the
increased probability of channeling events~\cite{Nastasi}.

The results we obtained for the W(110) and W(310) surfaces are summarized
in Tables~\ref{table4} and \ref{table5}. They are similar to the
results obtained for the W(100) surface (see Table~\ref{table1}) with a
few exceptions. For instance, the reflection coefficient for
deposition angles $30^{\circ}/60^{\circ}$ for the W(110) surface is
slightly lower/higher compared to the corresponding results for the
W(100) and W(310) surfaces. This can be explained by taking into
account the differences in crystallographic orientation of substrates
corresponding to different surfaces with respect to the deposition
directions defined by the deposition angles.

 We have also carried out additional MD simulations of He
  reflection and implantation with the W(100) surface under cumulative
  bombardment by the He atoms, which would be the case in a real
  fusion reactor.  Although the results we obtained are preliminary
  and further work is required, it is of interest to note that the values we obtained
  for the particle/energy reflection coefficients and the average
  implantation depth are very similar to those obtained using the clean W
  surface without any He implanted. We note again that the results
  reported in this paper correspond to non-cumulative
  simulations.

\subsection{Implantation depth or range distribution\label{subsec:results-range}}

\begin{figure}[t  ]
\includegraphics[width=7.0cm] {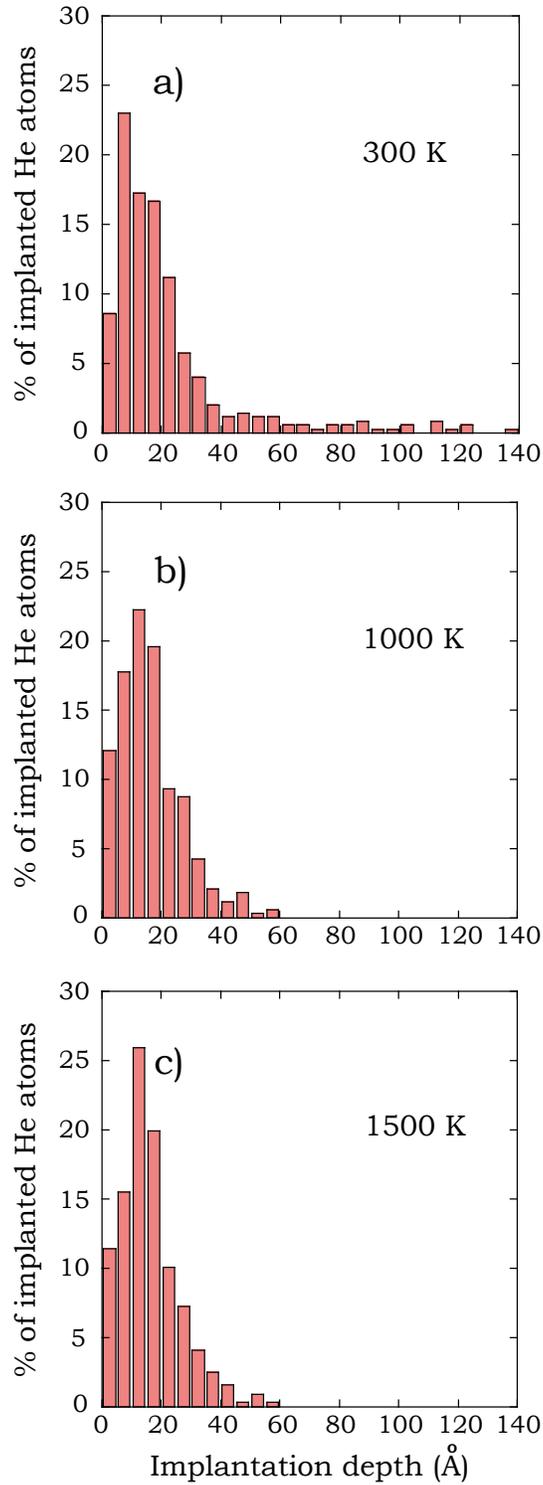}
\caption{\label{idd-temp} {Dependence of implantation depth
    distribution for He atoms on substrate temperature. All for normal incidence on
    W(100) surface. Initial energy of He atoms is
    80~eV. W(100) substrate temperature is: a) 300~K, b) 1000~K, c)
    1500~K. }}
\end{figure}

\begin{figure}[t  ]
\includegraphics[width=7.0cm] {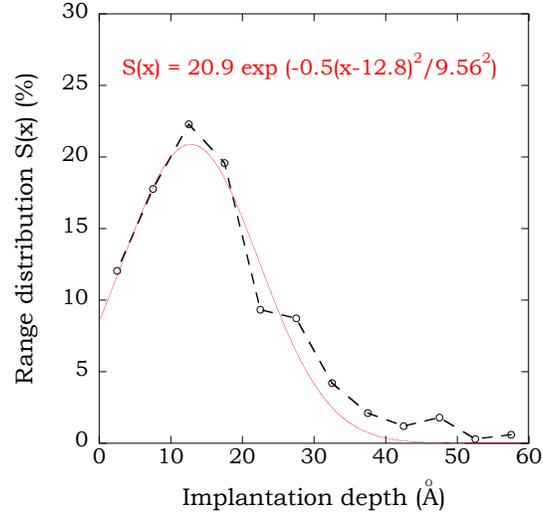}
\caption{\label{idd1000K-80ev-g} {The implantation
    depth distribution for the bulk of the low energy He atoms can be fitted by a 
Gaussian distribution truncated at $x=0$. This is for normal incidence on a W(100)
    surface. Initial energy of He atoms is 80~eV. The substrate
    temperature is $T = 1000$~K. }}
\end{figure}

For a plasma irradiation flux whose variation along the tungsten surface is small
over the length scale of the transverse spread of the implanted ions, 
the implantation quantity of practical utility is the angularly-integrated projected range function
\begin{widetext}
\begin{equation}
\bar{\cal S}(x | E_i,\theta_i) \equiv 
\int_0^{\pi/2} x^2 \tan\theta_s (1 + \tan^2\theta_s) d\theta_s
\int_0^{2\pi} d\varphi_s \int_0^{2\pi} d\varphi_i {\cal S}.
\end{equation}
\end{widetext}
Fig.~\ref{idd-temp} compares implantation depth distributions
$\bar{\cal S}(x|E_i,\theta_i)$ of He atoms for different substrate
temperatures (300~K, 1000~K and 1500~K). Normal deposition on W(100)
substrate is considered. Initial incident energy of He atoms is 80~eV
for all substrate temperatures. While the results are very similar for
T = 1000~K and T = 1500~K (see Fig.~\ref{idd-temp}b and
Fig.~\ref{idd-temp}c), in the case of T = 300~K, the implantation
depth distribution has a long tail due to a substantial increase in the
number of channeling events\hspace{1mm}\cite{Nastasi}.  The channeling
effect generally depends on substrate temperature, substrate
orientation with respect to a deposition direction and other
factors. We observed it in all cases corresponding to different
substrate temperatures, different W surfaces, different He incident
energies and angles. In the absence of the channeling effect the
implantation depth distributions are usually modeled by a
Gaussian distribution~\cite{Nastasi}:
\begin{equation}
  \bar{\cal S}(x) =
  \frac{\Phi_i}{\bigtriangleup R_p (2 \pi)^{\frac{1}{2}}} \exp [ -\frac{1}{2} (\frac{x - R_p}{\bigtriangleup R_p})^2 ].
\end{equation}
Where $R_p$ is known as the projected range and
$\bigtriangleup R_p$ is the projected range straggling (the average
fluctuation, or standard deviation from the mean in the projected
range).  $\Phi_i$ is a constant set by the implantation dose $N$ (or total number of
implanted He atoms)~\cite{Nastasi} such that
\begin{equation}
N =
\int_{0}^{\infty} \bar{\cal S}(x) dx.
\end{equation}
In the case of high energy ion implantation, $\Delta R_p \ll R_p$ can be satisfied, so
the lower bound of this integral can be set to $-\infty.$
The constant $\Phi_i$ is the implantation dose itself, {\em i.e.}, $\Phi_i = N.$
In the case of low energy implantation, as considered here,
$\Delta R_p$ is comparable to $R_p.$ This gives rise to a Gaussian distribution truncated 
at the small $x$ end, as shown in Fig.~\ref{idd1000K-80ev-g}, which provides a
 fit to the implantation
depth distribution of He atoms (1000~K, normal deposition on W(100)
surface, incident energy 80~eV).
The actual average projected range $L,$ as defined in Eq.~(\ref{eq:average-range}),
can be significantly different from $R_p$ in the Gaussian model. 

\begin{figure}[t  ]
\includegraphics[width=7.0cm] {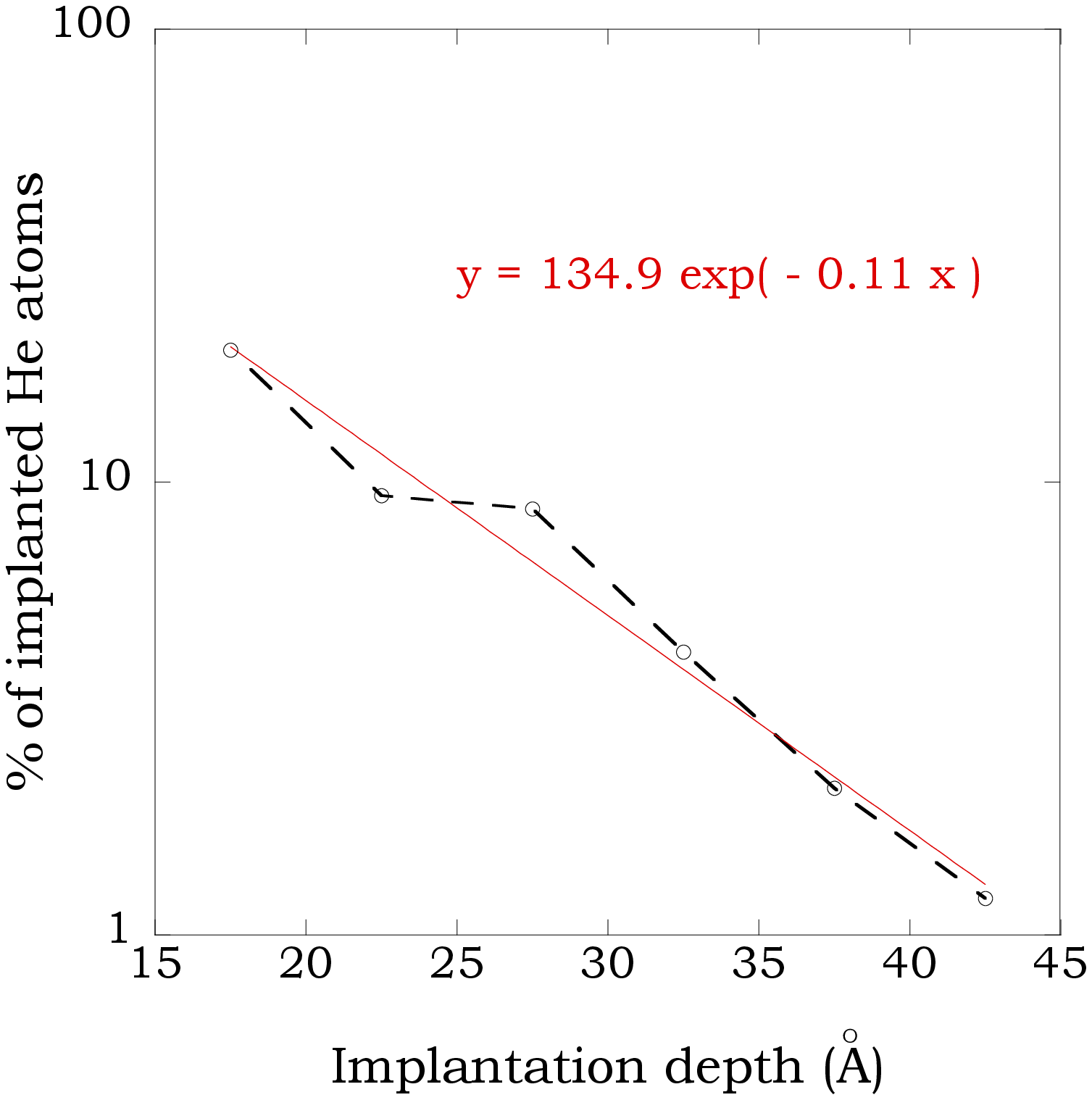}
\caption{\label{idd1000K-80ev-e} {The tail of the
    projected range distribution is exponential. Normal
    deposition. W(100) surface. Initial energy of He atoms is
    80~eV. The substrate temperature is $T = 1000$~K. }}
\end{figure}

\begin{figure}[t  ]
\includegraphics[width=7.0cm] {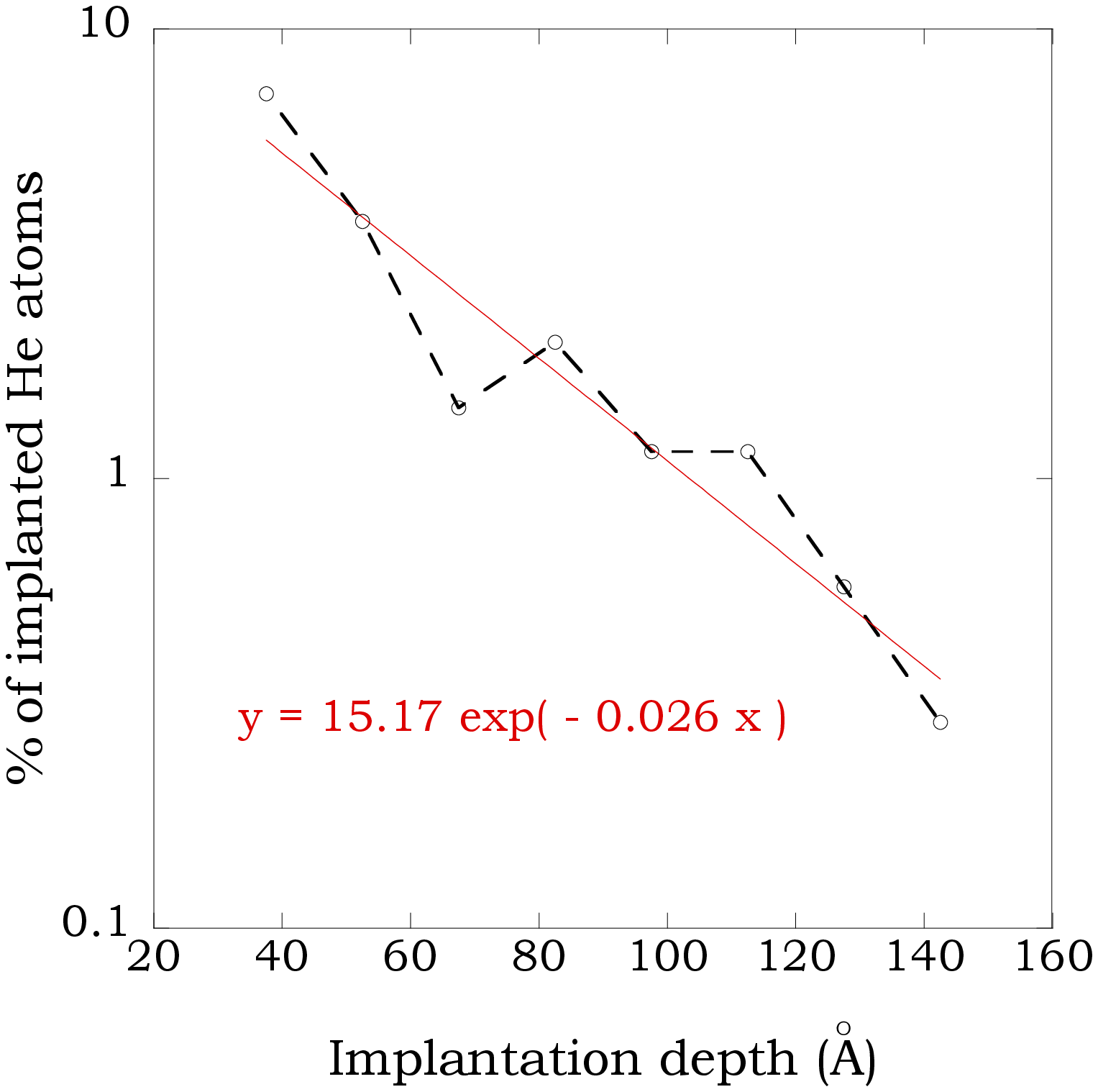}
\caption{\label{idd300K-80ev-e} {The tail of the
    projected range distribution is exponential. Normal
    deposition. W(100) surface. Initial energy of He atoms is
    80~eV. The substrate temperature is $T = 300$~K. }}
\end{figure}

The channeled He atoms that penetrate much beyond $R_p$ have a
distribution that falls off exponentially with distance as
$\exp(-x/\lambda_c)$, where $\lambda_c \gg R_p$ (see
Ref.~\onlinecite{Nastasi}).  Fig.~\ref{idd1000K-80ev-e} shows the
exponential fit to the tail of the same distribution in
Fig.~\ref{idd1000K-80ev-g}. In Fig.~\ref{idd300K-80ev-e} an exponential fit to the tail of
the implantation depth distribution is shown for simulation temperature $T=300$K 
(corresponds to the distribution from Fig.~\ref{idd-temp}a).
 In cases of higher temperatures (1000~K
and 1500~K), the maximum implantation depth does not exceed 60~\AA~and
the peaks of the distributions are between 10~\AA~and 15~\AA. On the
other hand, in the case of He implantation in W(100) at 300~K, the
maximum He implantation depth observed in our MD simulations is close
to 140~\AA. This can be explained by an increase in number of
channeling events at 300~K. At 300~K the peak of the distribution is
between 5~\AA~and 10~\AA.

\begin{figure}[t  ]
\includegraphics[width=7.0cm] {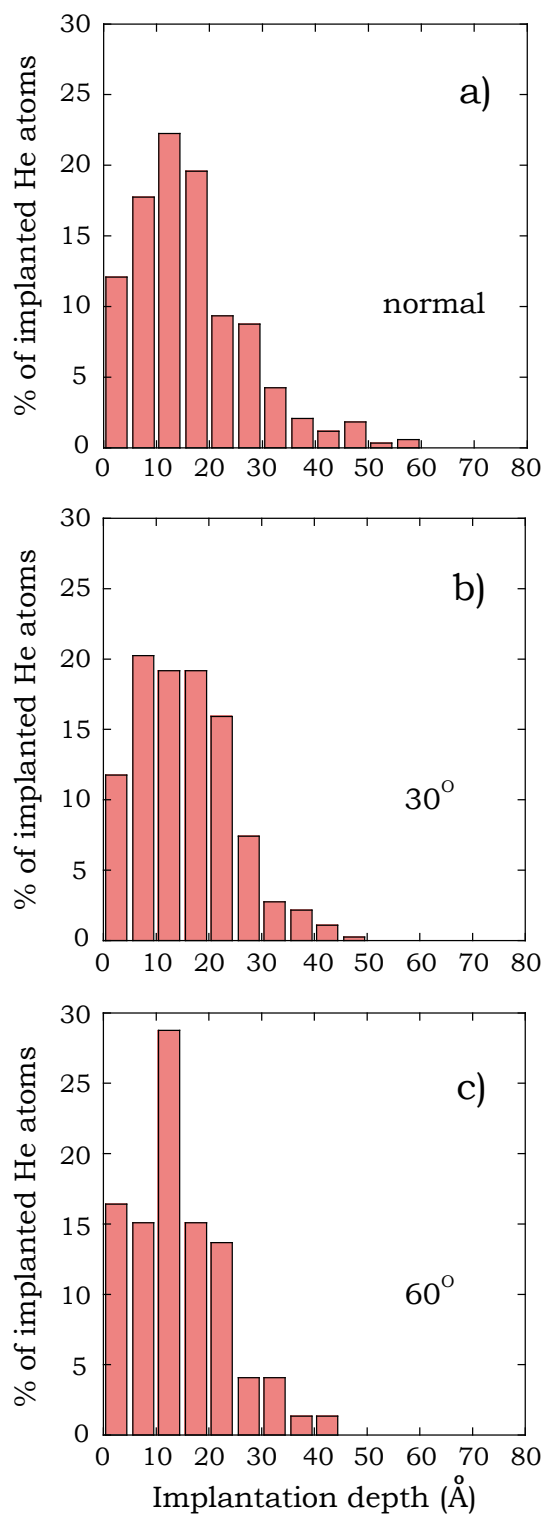}
\caption{\label{idd-angles} {Dependence of implantation depth
    distribution for He atoms on deposition angle. The substrate
    is W(100) at $T= 1000$~K. Initial energy of He atoms
    is 80~eV. The deposition angle is: a) $0^{\circ}$ (normal deposition),
    b) $30^{\circ}$, c) $60^{\circ}$.}}
\end{figure}

In Fig.~\ref{idd-angles} the dependence of implantation depth
distribution for He atoms on deposition angle is presented. Three
incidence angles with respect to substrate normal are considered
($0^{\circ}, 30^{\circ}, 60^{\circ}$). In all cases the temperature of the 
W(100) substrate is 1000~K and the incident energy of He atoms is
80~eV. Even though 1000 non-cumulative depositions were carried out for
each deposition angle, the number of implantation events is not enough
to accumulate good statistics for the large incidence angle of
$60^{\circ}$ (the number of implanted He atoms is only 73, as can be
seen from Table~\ref{table1}). We can see from Fig.~\ref{idd-angles}a
that in the case of normal deposition, the projected range
distribution has a slightly longer tail, compared to
Figs.~\ref{idd-angles}b and \ref{idd-angles}c, which correspond to
the deposition angles of $30^{\circ}$ and $60^{\circ},$
respectively. 
This is due to the increased number of channeling events 
expected for the deposition angle of $0^{\circ}$ (and $45^{\circ}$) 
with respect to substrate normal in the case of the W(100) surface.
As a result, the average implantation depth is also
slightly increased in the case of normal deposition (see
Table~\ref{table1}).

\begin{figure}[t  ]
\includegraphics[width=7.0cm] {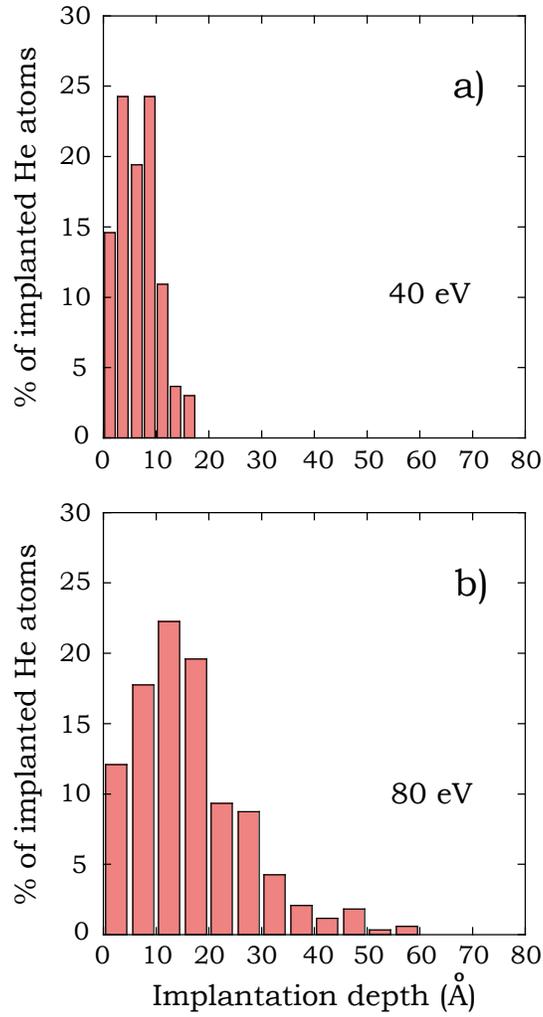}
\caption{\label{idd-enrg} Dependence of implantation depth
    distribution for He atoms on the initial energy of He atoms. This is for normal
    deposition on a W(100)  substrate at $T= 1000$~K Initial energy of He atoms is: a) 40~eV, b) 80~eV. Bin
    size is 2.5~\AA~and 5~\AA~respectively.}
\end{figure}

Fig.~\ref{idd-enrg} demonstrates the dependence of implantation depth
distribution for He atoms on the incident energy (incident energies of
40~eV and 80~eV are considered). In both
cases deposition is normal and the temperature of the W(100) substrate is
1000~K. As can be seen from Fig.~\ref{idd-enrg}, initial
energy of the He atom has a strong effect on the implantation depth
distribution. He atoms with initial kinetic energy of
40~eV do not propagate deeper than
17.5~\AA~into the W substrate. On the other hand, the
increase of incident energy to 80~eV results in a
substantial increase in average implantation depth for He atoms (see
Table~\ref{table1}). The maximum propagation distance increases up to
60~\AA~in this case.

\begin{figure}[t  ]
\includegraphics[width=7.0cm] {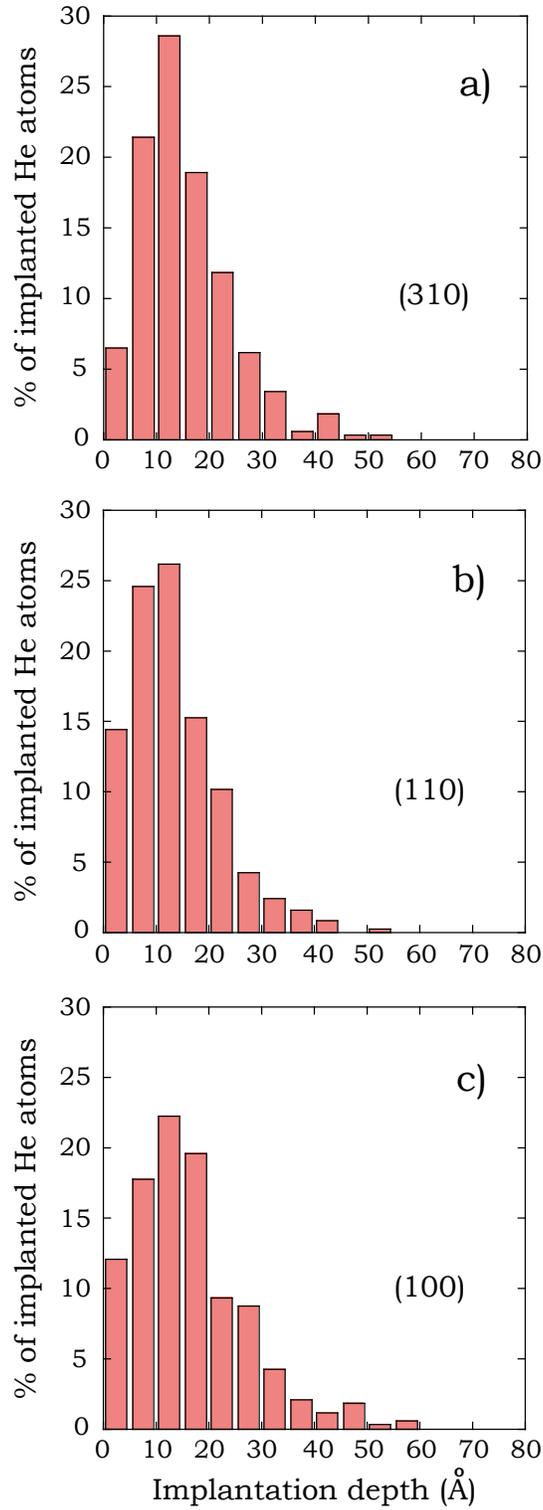}
\caption{\label{idd-surfaces} {Dependence of implantation depth
    distribution for He atoms on substrate surface type is weak for
    fusion relevant wall temperature. This is for normal deposition
    and W substrate temperature is $T= 1000$~K. Initial energy of He
    atoms is 80~eV. W substrate surface is: a) (310), b) (110), c)
    (100).}}
\end{figure}

The implantation depth distributions for He atoms for the W(100), W(110)
and W(310) surfaces are presented in Fig.~\ref{idd-surfaces}.  In all
three cases the temperature of the W substrate is 1000~K and the
incident energy of normally deposited He atoms is 80~eV.  As can
  be seen from Fig.~\ref{idd-surfaces}, the implantation profiles
  follow a similar behavior, but different lattice orientations do
  produce quantitative variations. We also note that all three surfaces show the 
limited implantation depth arising from disruption of channeling at this high, fusion-relevant
temperature of T=1000K, consistent with what was seen in Fig.~\ref{idd-temp}.

\subsection{Energy distribution of He atoms reflected by W surface\label{subsec:results-energy}}

The energy reflection coefficient $R_E$ is of importance to boundary
plasmas since it is a measure of how efficiently plasma ion energy is
being transferred into the reactor wall/divertor.
The reflected ion energy distribution $\left<F_r\right>_{\theta_r,\varphi_r}$
is 
\begin{widetext}
\begin{equation}
\bar{F}_r(E_r | E_i,\theta_i, \varphi_i; \Sigma, T) = \left<F_r\right>_{\theta_r,\varphi_r}
=
\int_0^{\pi/2} \sin\theta_r d\theta_r \int_0^{2\pi} d\varphi_r {\cal T}. 
\end{equation}
\end{widetext} 

\begin{figure}[t  ]
\includegraphics[width=7.0cm] {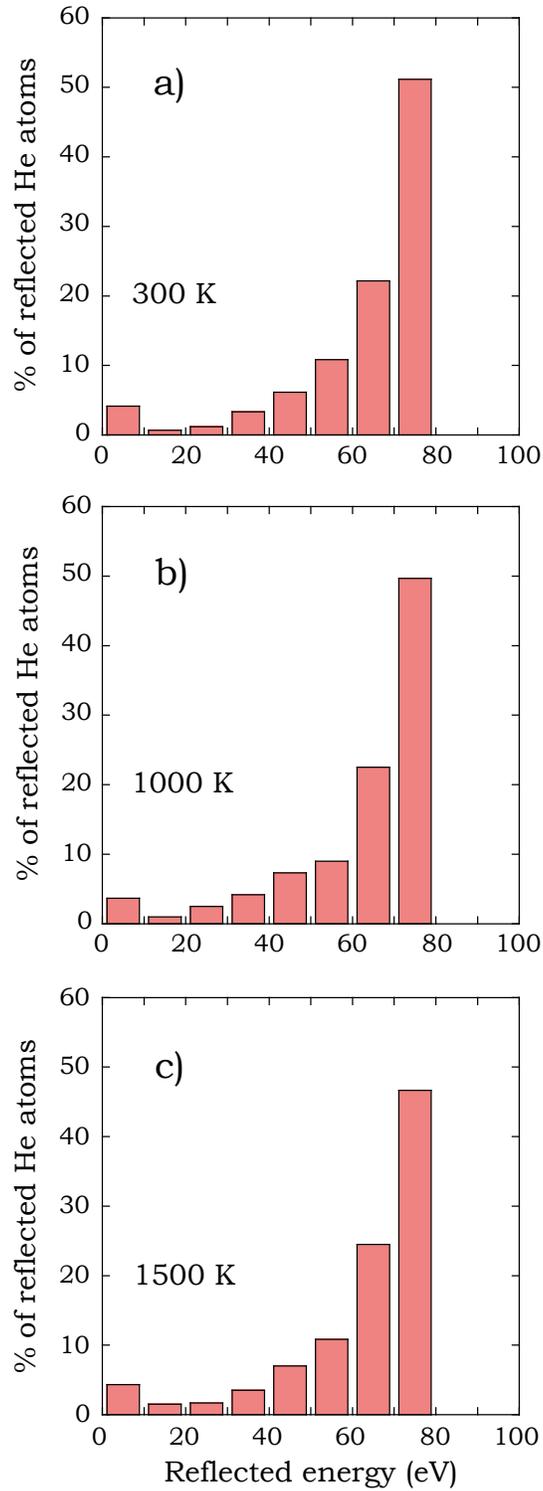}
\caption{\label{red-temp} {Dependence of reflection energy
    distribution for He atoms on substrate temperature. This is for normal
    deposition on a W(100) surface. Initial energy of He atoms is
    80~eV. W(100) substrate temperature is: a) 300~K, b) 1000~K, c)
    1500~K. }}
\end{figure}

In Fig.~\ref{red-temp} we compare energy distributions of reflected
He atoms for different substrate temperatures (300~K,
1000~K and 1500~K). In all cases He atoms
were deposited normally onto the W(100) substrate. The incident energy of He
atoms is 80~eV for all substrate temperatures. As can be
seen from Fig.~\ref{red-temp}, the substrate temperature does not
affect the results significantly.

\begin{figure}[t  ]
\includegraphics[width=7.0cm] {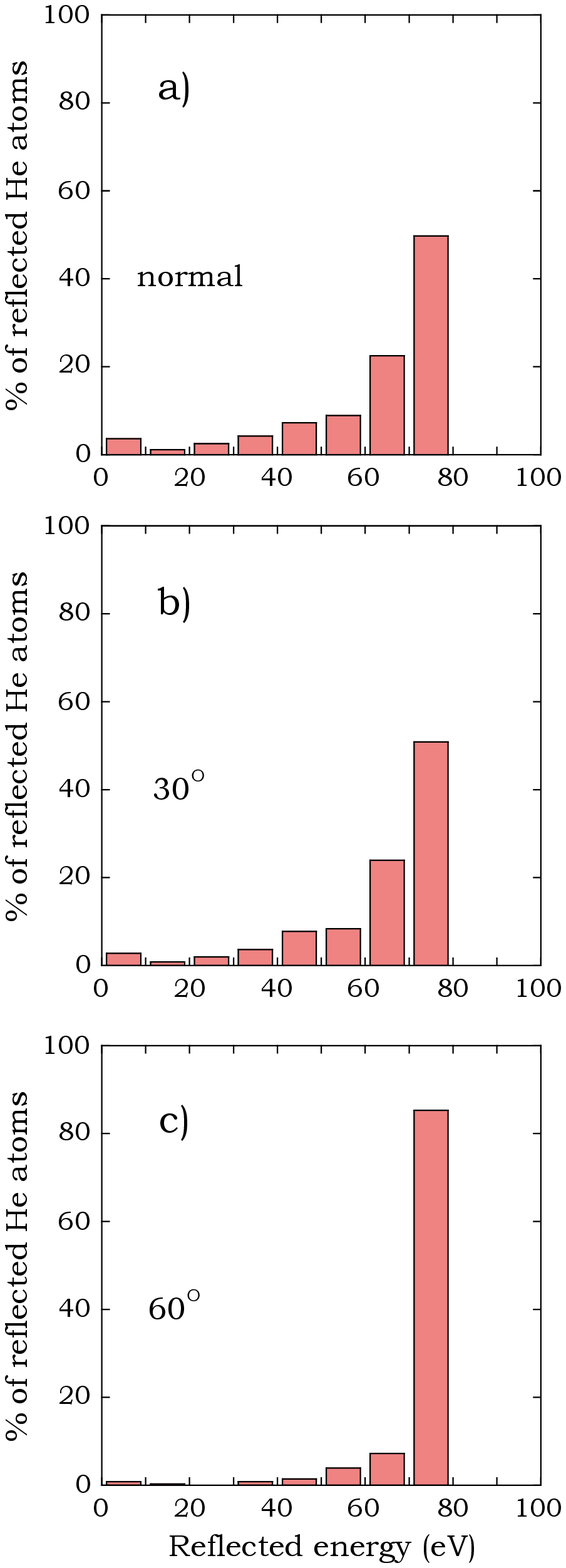}
\caption{\label{red-angles} {Dependence of reflection energy
    distribution for He atoms on deposition angle. The substrate
    is W(100) at $T= 1000$~K. Initial energy of He atoms
    is 80~eV. Deposition angle is: a) $0^{\circ}$ (normal deposition),
    b) $30^{\circ}$, c) $60^{\circ}$.}}
\end{figure}

Fig.~\ref{red-angles} shows the dependence of reflected He energy
distribution on deposition angle. As can be seen from
Fig.~\ref{red-angles}, the dependence is very strong. For lower
deposition angles ($0^{\circ}, 30^{\circ}$) about 50\%
of reflected He atoms have kinetic energy close to the initial
incident energy. For the $60^{\circ}$ deposition angle the percentage of
He atoms reflected with kinetic energy 70~eV -
80~eV is close to 90. This can be explained by the fact
that at higher deposition angles a probability of substrate
penetration by He atoms is reduced significantly. As a result, the
fraction of He atoms that have a chance to lose energy in collisions
with the atoms of W substrate before they find a way out is
decreased substantially compared to the cases of lower deposition
angles.

\begin{figure}[t  ]
\includegraphics[width=7.0cm] {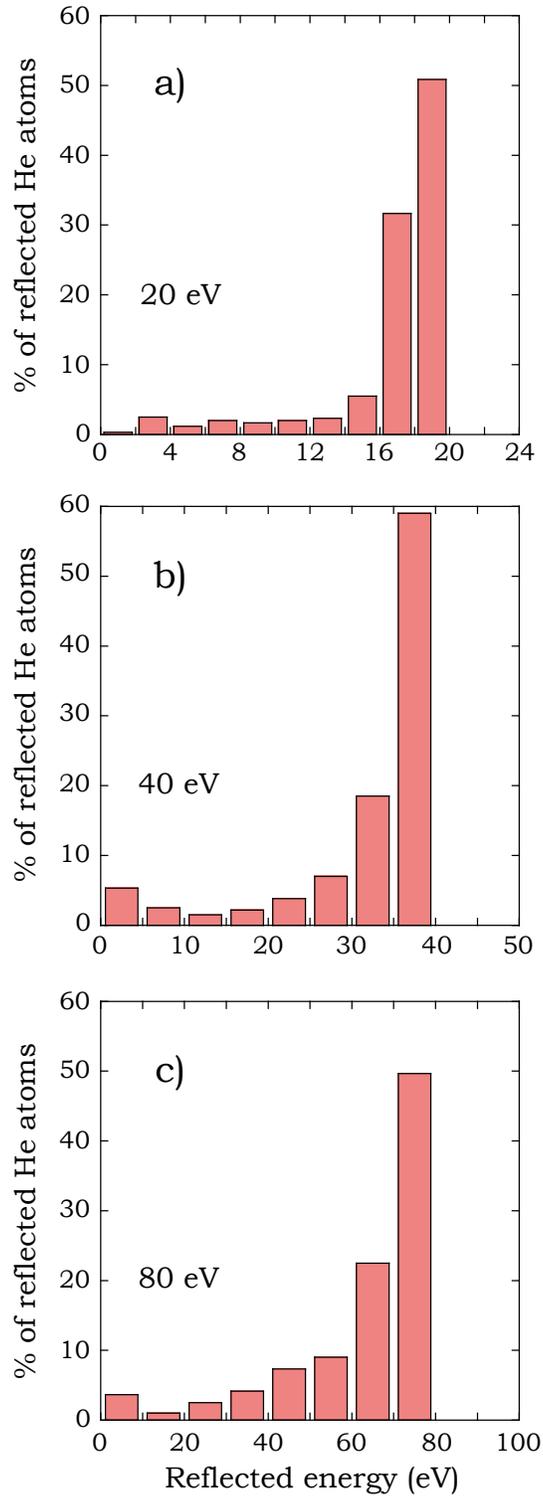}
\caption{\label{red-enrg} {Dependence of reflection energy
    distribution for He atoms on the incidence energy. This is for normal
    deposition on a W(100) surface. W substrate temperature is
    1000~K. Initial energy of He atoms is: a) 20~eV, b) 40~eV, c)
    80~eV. Energy bin size is 2~eV, 5~eV and 10~eV respectively.}}
\end{figure}

\begin{figure}[t  ]
\includegraphics[width=7.0cm] {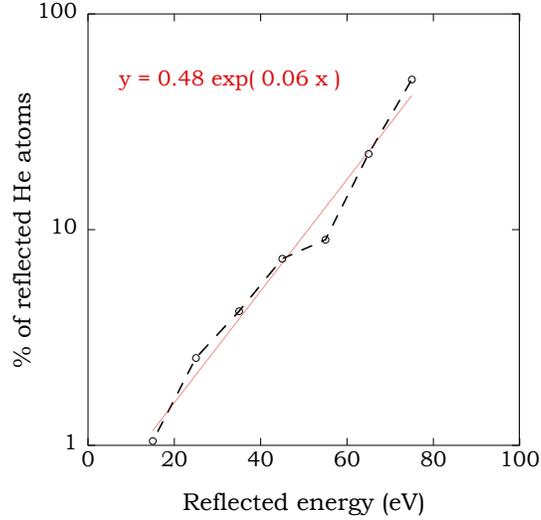}
\caption{\label{red1000K-80ev-e} {In the case of 80~eV incidence He energy,
  the reflection energy distribution for He atoms (from 20~eV to 80~eV) follows an exponential
  distribution of $E_r$. This is for normal
  deposition on a W(100) surface. W(100) substrate temperature is
  $T= 1000$~K. The energy bin size is 10~eV.}}
\end{figure}

\begin{figure}[t  ]
\includegraphics[width=7.0cm] {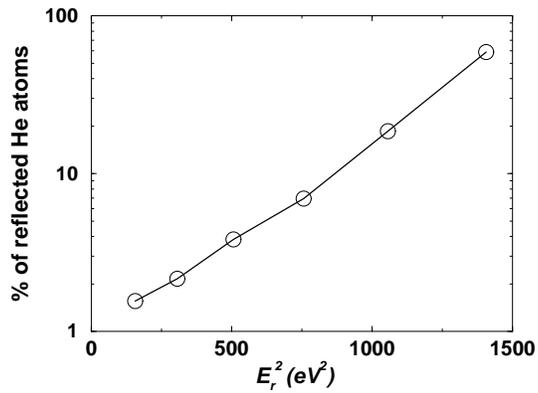}
\caption{\label{red1000K-40ev-g} { In the case of 40~eV incidence He
    energy, the reflection energy distribution for He atoms (from
    15~eV to 40~eV) follows a Gaussian tail distribution in $E_r$
    instead, in contrast with the $E_i=80$~eV case
    (Fig.~\ref{red1000K-80ev-e}). This is for normal deposition on a
    W(100) surface. W(100) substrate temperature is $T= 1000$~K. Energy bin
    size is 5~eV.}}
\end{figure}

In Fig.~\ref{red-enrg} the dependence of reflected He energy
distribution on incident energy is shown. Incident energies of 20~eV,
40~eV and 80~eV are considered. In all cases the temperature of the W(100)
substrate is 1000~K, and He atoms are deposited normally with respect
to the substrate surface. As can be seen from Fig.~\ref{red-enrg}, the
results are not very different in all three cases. However, the
scaling of the results with $E_r$ is not the same.  As can be seen from
Fig.~\ref{red1000K-80ev-e}, the reflected energy distributions
in the  higher He incidence energy case ($E_i=80$~eV is shown) closely follow an exponential
distribution in $E_r$ (up to $E_i$), 
\begin{equation}
\label{eq:reflect-exp}
\bar{F}_r \sim \exp(\alpha E_r).
\end{equation}
In contrast, the reflected energy
distribution of the low incidence energy case ($E_i=40$~eV is shown) approximately scales as
$\exp(\beta E_r^2)$ (see Fig.~\ref{red1000K-40ev-g}),
\begin{equation}
\label{eq:reflect-gaussian}
\bar{F}_r \sim \exp(\beta E_r^2).
\end{equation}
How this qualitative transition in $E_r$ scaling of $\bar{F}_r$ occurs
as $E_i$ increases remains to be understood from a theoretical perspective.

\begin{figure}[t  ]
\includegraphics[width=7.0cm] {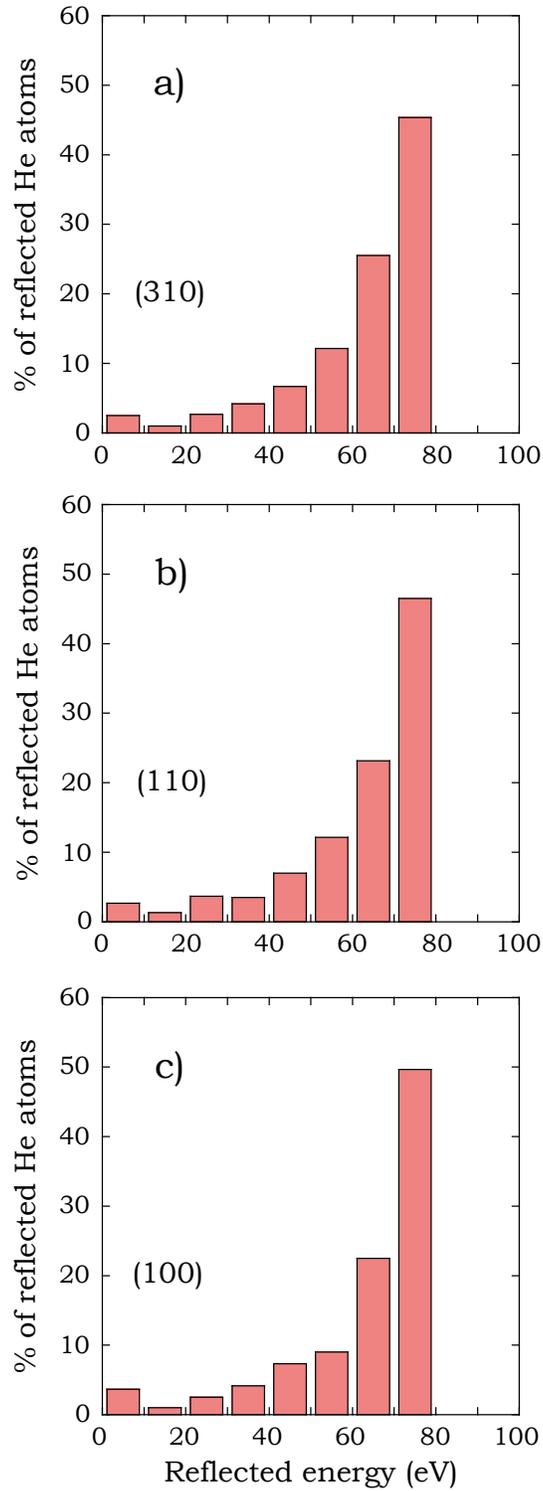}
\caption{\label{red-surfaces} {Dependence of reflection energy
    distribution for He atoms on substrate surface type. This is for normal
    deposition. W substrate temperature is $T= 1000$~K. Initial energy of
    He atoms is 80~eV. W substrate surface is: a) (310), b) (110), c)
    (100).}}
\end{figure}

The dependence of the reflected energy distribution for He atoms on the
type of W surface (W(100), W(110) and W(310) surfaces are considered)
is also weak (see Fig.~\ref{red-surfaces}).  In all cases considered
in Fig.~\ref{red-surfaces}, the temperature of the W substrate is 1000~K
and the incident energy of He atoms is 80~eV.

\subsection{Angular distribution  of He reflection by W surface\label{subsec:results-angular}}

\begin{figure}[t  ]
\includegraphics[width=7.0cm] {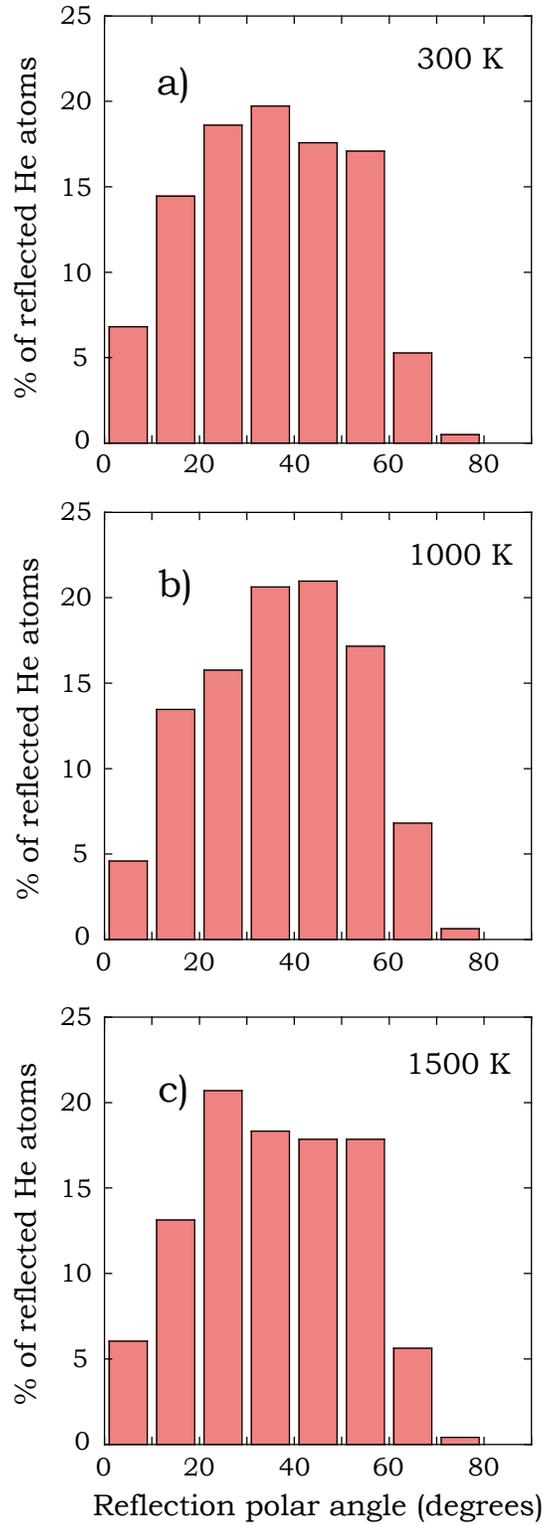}
\caption{\label{polar-temp} {Dependence of distribution of polar
    angles of reflection for He atoms on substrate temperature. These are for normal
    deposition on W(100) surfaces. Initial energy of He atoms is
    80~eV. W(100) substrate temperature is: a) 300~K, b) 1000~K, c)
    1500~K. }}
\end{figure}

\begin{figure}[t  ]
\includegraphics[width=7.0cm] {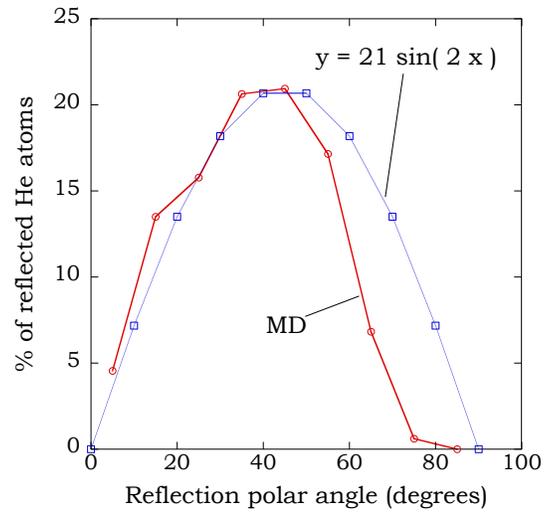}
\caption{\label{cosine} {A sine fit to the distribution of polar
    angles of reflection for He atoms. MD results correspond to normal
    deposition on W(100) surface. Initial energy of He atoms is
    80~eV. W(100) substrate temperature is $T= 1000$~K. }}
\end{figure}

Fig.~\ref{polar-temp} shows the dependence of the polar angle distribution
of reflected He atoms on substrate temperature (300~K, 1000~K and
1500~K). In all three cases the incident energy of He atoms on the W(100)
substrate is 80~eV, and deposition is normal to the substrate
surface. The same trend is observed in that most of the He
atoms have the polar angle of reflection in the interval of $10^{\circ}
- 60^{\circ}$ with respect to substrate normal. It was indicated in
previous studies on this topic that the distribution of polar angles
of reflection for normal incidence has a characteristic
sine distribution in $\theta_r$~\cite{Eckstein2}. In our data, the
polar angle distributions of the reflected He atoms
deviate from the sine distribution in such a way that more He atoms
are reflected at small polar angles while fewer He atoms
are reflected at large polar angles (for example, see Fig.~\ref{cosine}).

\begin{figure}[t  ]
\includegraphics[width=7.0cm] {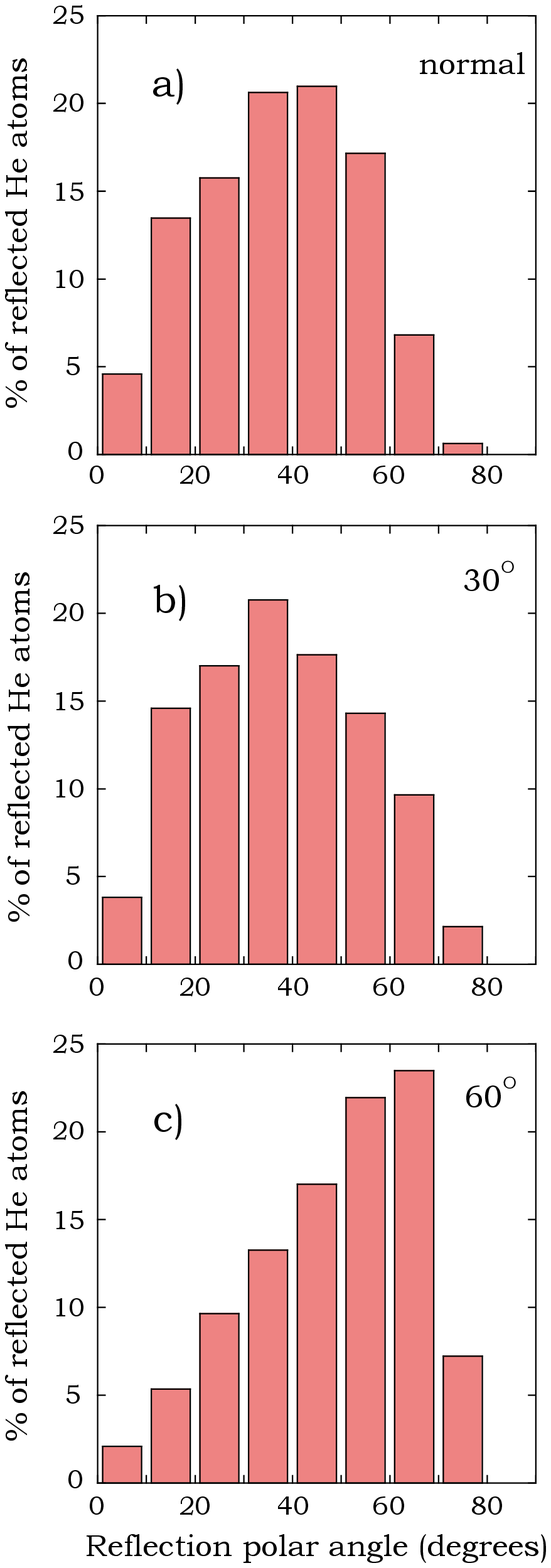}
\caption{\label{polar-angles} {Dependence of distribution of polar
    angles of reflection for He atoms on deposition angle. The substrate
    is W(100) at $T= 1000$~K. Initial energy of He atoms
    is 80~eV. Deposition angle is: a) $0^{\circ}$ (normal deposition),
    b) $30^{\circ}$, c) $60^{\circ}$.}}
\end{figure}

In Fig.~\ref{polar-angles} the dependence of polar angle distribution
of reflected He atoms on deposition (polar) angle is shown. From
Fig.~\ref{polar-angles}b and Fig.~\ref{polar-angles}c, showing 
incident angles of $30^{\circ}$ and $60^{\circ}$, one can see
that the peak of the distribution is very close to the initial polar
angle of deposition (the azimuthal angle is of course opposite). This
is especially pronounced for higher angles of incidence. In these
cases He atoms do not penetrate the W substrate very often, and, as a
result, the polar angle of reflection is very close to the polar angle
of incidence. At $\theta = 0^{\circ}$ (Fig.~\ref{polar-angles}a), the 
distribution peak is not at all aligned with the incoming angle, because 
in this case the He atoms penetrate deeper (and more often) into the substrate, 
and spend some time under the surface colliding with the atoms of the substrate, 
and hence lose their memory of the incoming angle. 

\begin{figure}[t  ]
\includegraphics[width=7.0cm] {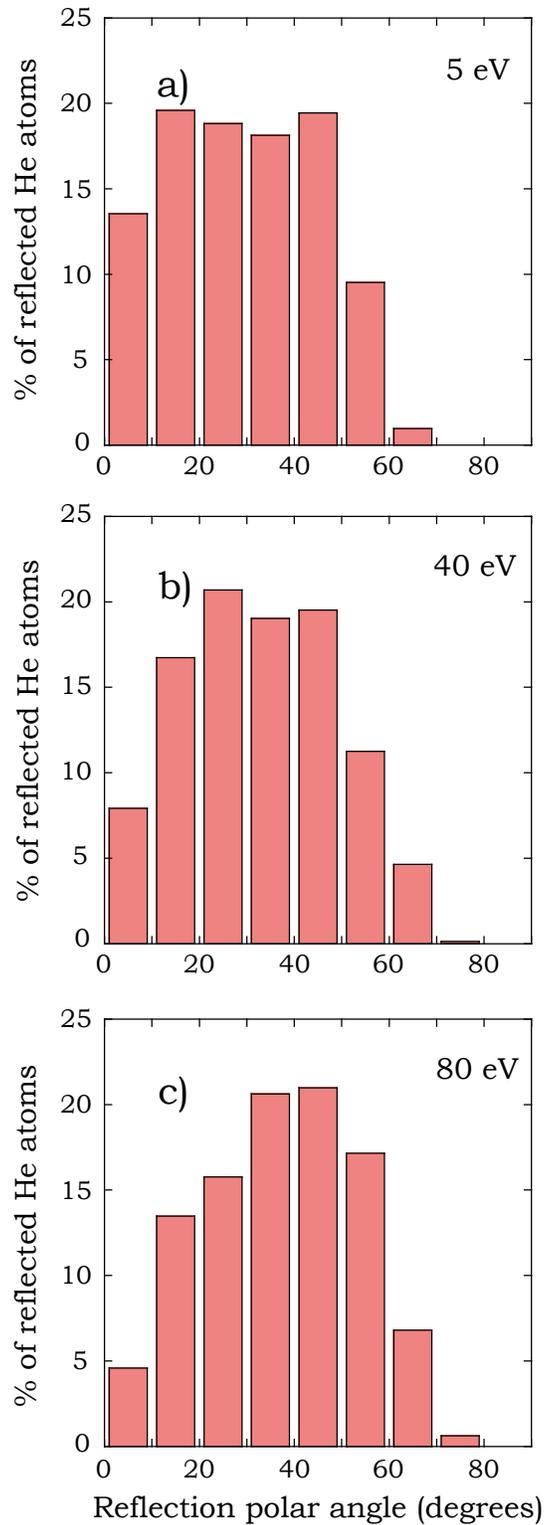}
\caption{\label{polar-enrg} {Dependence of distribution of polar
    angles of reflection for He atoms on incidence energy of He
    atoms. These are for normal deposition on W(100) surfaces. The W substrate
    temperature is 1000~K. Incidence energy of He ions is: a) 5~eV, b)
    40~eV, c) 80~eV.}}
\end{figure}

Fig.~\ref{polar-enrg} illustrates the dependence of the polar
angle distribution of reflected He atoms on the incident energy
(incident energies of 5~eV 40~eV and
80~eV are considered). In all cases He atoms are
deposited normally on the W(100) surface, and the temperature of the W substrate
is 1000~K. The distribution has a more
pronounced peak in the case of higher He incident energy (see
Fig.~\ref{polar-enrg}c).

\begin{figure}[t  ]
\includegraphics[width=7.0cm] {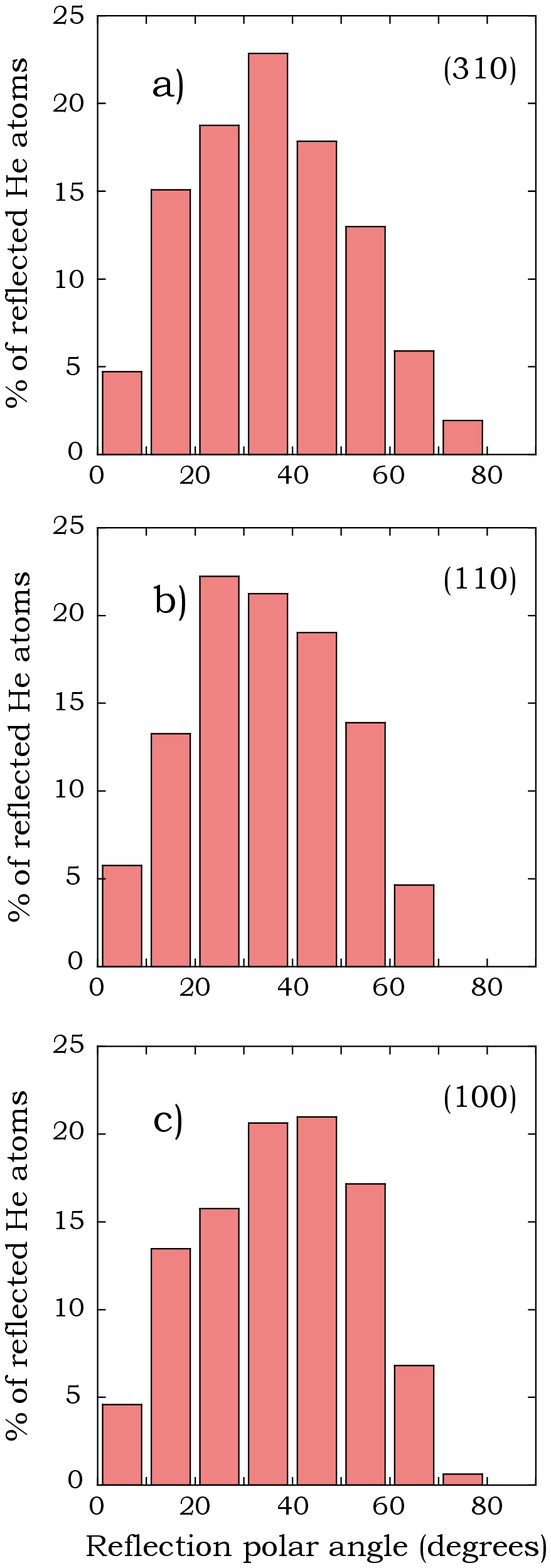}
\caption{\label{polar-surfaces} {Dependence of distribution of polar
    angles of reflection for He atoms on substrate surfaces. These are for normal
    deposition. W substrate temperature is  $T= 1000$~K. Incidence energy of
    He atoms is 80~eV. W substrate surface is: a) (310), b) (110), c)
    (100).}}
\end{figure}

We do not observe a strong dependence of polar angle distribution of
the reflected He atoms on the type of W surface (see
Fig.~\ref{polar-surfaces}). There are three surfaces considered: W(100), W(110) and W(310).
In all cases the temperature of the W substrate is 1000~K and
the incident energy of He atoms is 80~eV.

\begin{figure}[t  ]
\includegraphics[width=7.0cm] {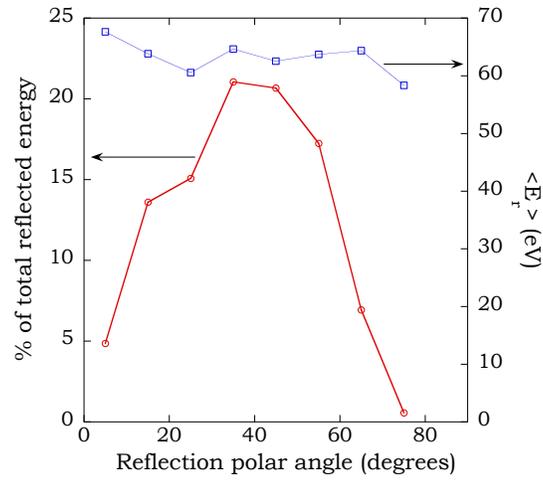}
\caption{\label{polar-vs-kinenrg} {Dependence of reflection energy
    distribution for He atoms on the reflection polar angle. And the dependence 
    of average kinetic energy reflected at a particular polar angle on the reflection polar angle. This is for normal
    deposition on a W(100) surface. Incidence energy of He atoms is
    80~eV. W substrate temperature is  $T= 1000$~K. }}
\end{figure}

One can also integrate the total reflected He energy 
over $E_r$ at fixed $\theta_r.$  A representative result for the reflection energy
as a function of the reflection polar angle is
shown in Fig.~\ref{polar-vs-kinenrg}. 
Also shown in Fig.~\ref{polar-vs-kinenrg} is an average kinetic energy 
reflected at a particular polar angle as a function of reflection polar angle. 
We can see that it is roughly the same and independent 
of polar angle of reflection $\theta_r$.

\begin{figure}[t  ]
\includegraphics[width=7.0cm] {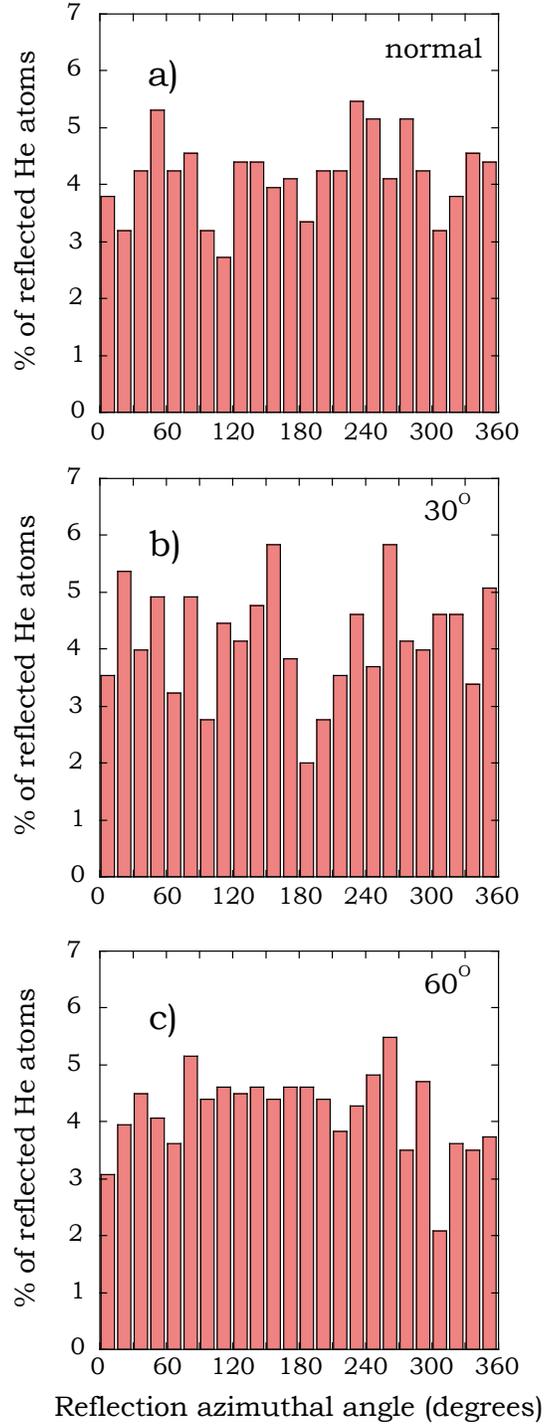}
\caption{\label{azimuthal-angles} {Dependence of distribution of
    azimuthal angles of reflection for He atoms on incidence (polar)
    angle. The substrate
    is W(100) at $T= 1000$~K. Incidence
    energy of He atoms is 80~eV. Deposition angle is: a) $0^{\circ}$
    (normal deposition), b) $30^{\circ}$, c) $60^{\circ}$.}}
\end{figure}

\begin{figure}[t  ]
\includegraphics[width=7.0cm] {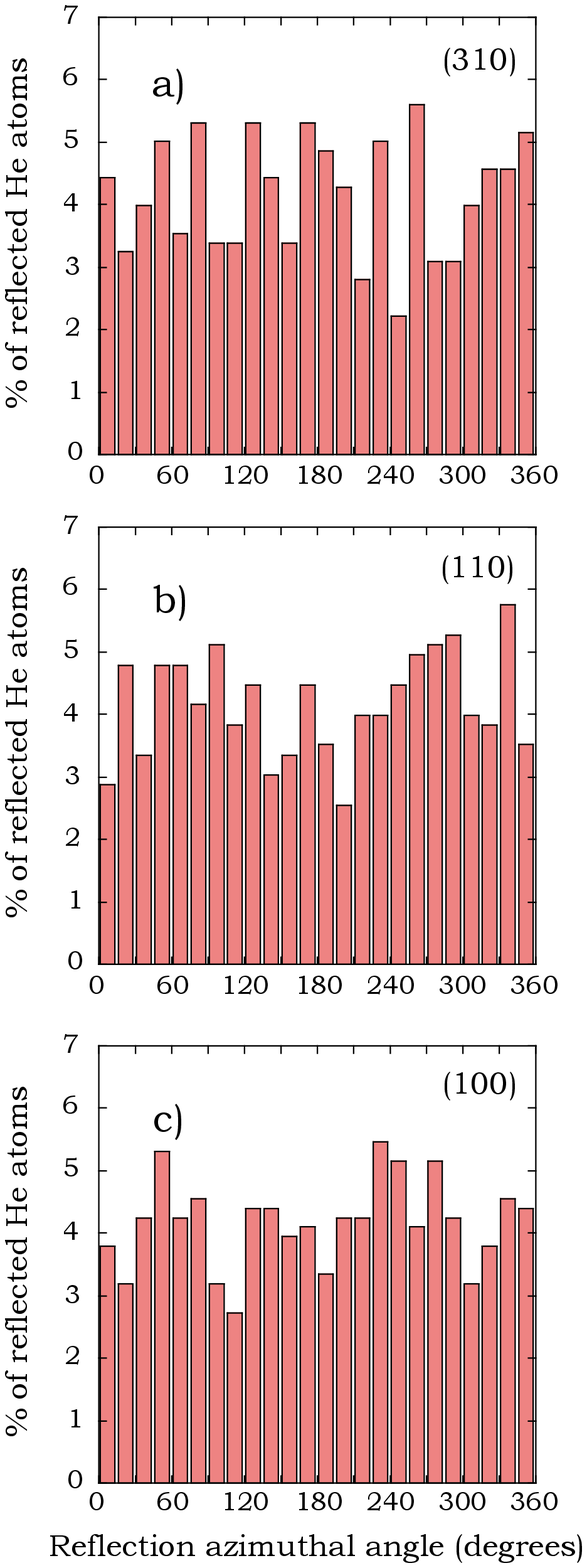}
\caption{\label{azimuthal-surfaces} {Dependence of distribution of
    azimuthal angles of reflection for He atoms on substrate
    surfaces. These are for normal deposition. W substrate temperature is
    $T= 1000$~K. Incidence energy of He atoms is 80~eV. W substrate surface
    is: a) (310), b) (110), c) (100).}}
\end{figure}

We could not detect any clear trend in the dependence of azimuthal
angle distribution of reflected He atoms on MD simulation parameters,
such as substrate temperature, deposition (polar) angle of He atoms,
incident energy of He atoms, or the type of W surface. It appears that the
azimuthal angles of reflection are distributed more or less evenly
over $360^{\circ}$. The dependence of azimuthal angle distribution
of reflection on deposition (polar) angle is shown in
Fig.~\ref{azimuthal-angles} ($0^{\circ}, 30^{\circ}$ and $60^{\circ}$
polar deposition angles are considered). We note that in all simulation cases of
non-normal deposition, the azimuthal
angles of incidence were chosen randomly. The distributions of
azimuthal angles of reflection for He atoms for W(100), W(110) and
W(310) surfaces are presented in Fig.~\ref{azimuthal-surfaces}.

\subsection{Results obtained using SRIM simulation package\hspace{1mm}\cite{SRIM}\label{subsec:results-srim}}

In Table~\ref{table6} the results of SRIM simulation~\cite{SRIM} are
presented. Normal incidence of He atoms (incidence energies ranging
from 5~eV to 100~eV) on
10000~\AA~thick W substrate is studied. Comparing the
results of Table~\ref{table6} with the corresponding results (normal
incidence) of Table~\ref{table1} one can clearly see a strong
disagreement between the SRIM and MD simulation results. SRIM
significantly underestimates the reflection coefficients for the whole
range of incidence energies considered in this study. The disagreement
is most pronounced for low incident energies.  The average
kinetic energy of reflected He atoms and the average implantation depth of
implanted He atoms, obtained using SRIM, are also underestimated,
compared to MD results. The disagreement between SRIM and
MD simulation results can be explained by taking into account that
SRIM uses a binary collision approximation to describe interactions
between an incident He atom and the atoms of the W substrate. In contrast, 
MD simulation takes into account many-body
interactions, which become increasingly important at low incident
energies.

\subsection{Results obtained using EAM potential for W by Derlet {\it et al}.\hspace{1mm}\cite{Derlet}\label{subsec:results-derlet}}

\begin{figure}[t  ]
\includegraphics[width=7.0cm] {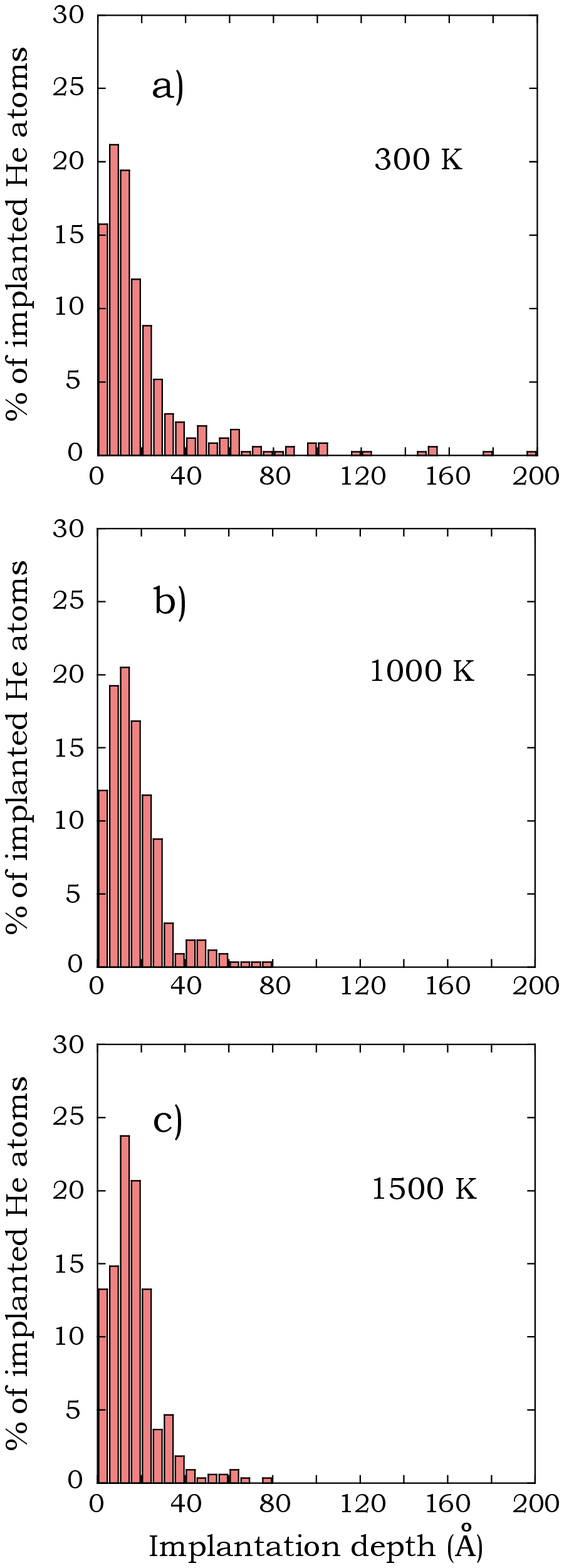}
\caption{\label{idd-temp-d} {EAM potential for W by Derlet {\it et
      al}~\cite{Derlet}. Dependence of implantation depth
    distribution for He atoms on substrate temperature. These are for normal
    deposition on a W(100) surface. Incidence energy of He atoms is
    80~eV. W(100) substrate temperature is: a) 300~K, b) 1000~K, c)
    1500~K. }}
\end{figure}

\begin{figure}[t  ]
\includegraphics[width=7.0cm] {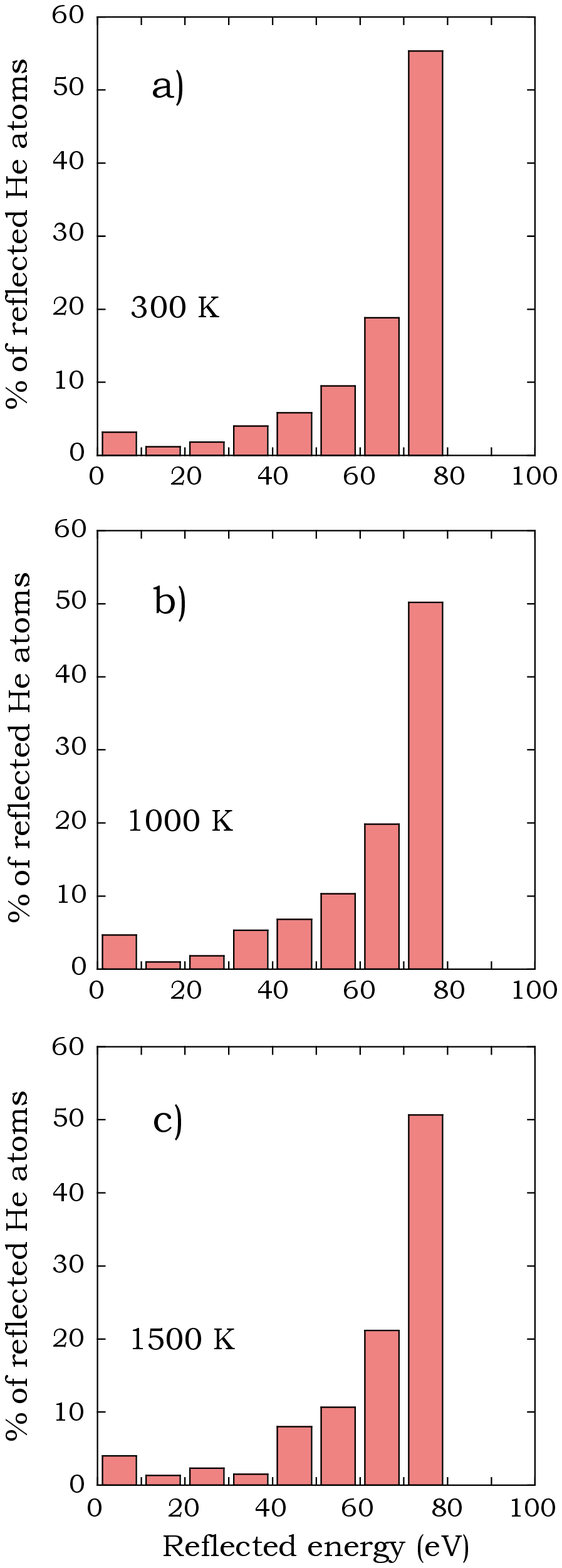}
\caption{\label{red-temp-d} {EAM potential for W by Derlet {\it et
      al}.~\cite{Derlet}. Dependence of reflection energy distribution
    for He atoms on substrate temperature. These are for normal deposition on a W(100)
    surface. Incidence energy of He atoms is 80~eV. W(100) substrate
    temperature is: a) 300~K, b) 1000~K, c) 1500~K. }}
\end{figure}

We carried out additional MD simulations of interaction of low energy
He atoms with the W(100) substrate using a different EAM interatomic
potential for W developed by Derlet {\it et
  al}.\hspace{1mm}\cite{Derlet} and modified by Bj\"{o}rkas et
al.\cite{Bjorkas, Bjorkas2}, while maintaining the same W-He cross potential
developed by Juslin {\it et al}~\cite{Juslin}.  
Comparing the results obtained using
this EAM potential for W (see
Figs.~\ref{idd-temp-d},\hspace{0.4mm}\ref{red-temp-d}) with the
corresponding results obtained using the Ackland-Thetford EAM potential
for W,\hspace{1mm}\cite{Ackland} modified by Juslin {\it et
  al}.\hspace{1mm}\cite{Juslin} (see
Figs.~\ref{idd-temp},\hspace{0.4mm}\ref{red-temp}), we can see that
they are not very different from each other. In particular, the
implantation depth distributions obtained for three different
substrate temperatures (300~K, 1000~K and
1500~K) using the Derlet potential (see Fig.~\ref{idd-temp-d}),
have longer tails (due to channeling events), compared to the
corresponding results obtained using the Ackland-Thetford potential (Fig.~\ref{idd-temp}). This is
especially pronounced at low temperature (300~K). At this
temperature the maximum He implantation depth observed in our MD
simulations (using the Derlet  potential) is close to 200~\AA.

The reflection energy distributions for He atoms for different
substrate temperatures (300~K, 1000~K and
1500~K) obtained using the Derlet  potential (see
Fig.~\ref{red-temp-d}) look very similar to the corresponding
distributions obtained using the Ackland-Thetford potential 
(see Fig.~\ref{red-temp}). In the former
case slightly more He atoms are reflected with the energy
close to the incident energy (80~eV).

\section{Comparison with existing results\label{sec:comparison}}

There has been a fairly limited number of experiments on low energy He
interaction with W surfaces. Among them are the experiments by van
Gorkum {\it et al}. (see Ref.~\onlinecite{vanGorkum}) and the
experiments by Amano {\it et al}. (see Ref.~\onlinecite{Amano}). The
process of interaction of low energy He with W surfaces was simulated
using the computer program MARLOWE (Version 11.5) (see
Ref.~\onlinecite{Robinson}). There are also a number of empirical
formulas (see Refs.~\onlinecite{Tabata, Eckstein, Thomas}) which were
fitted using experimental and simulation data.

\begin{figure}[t  ]
\includegraphics[width=7.0cm] {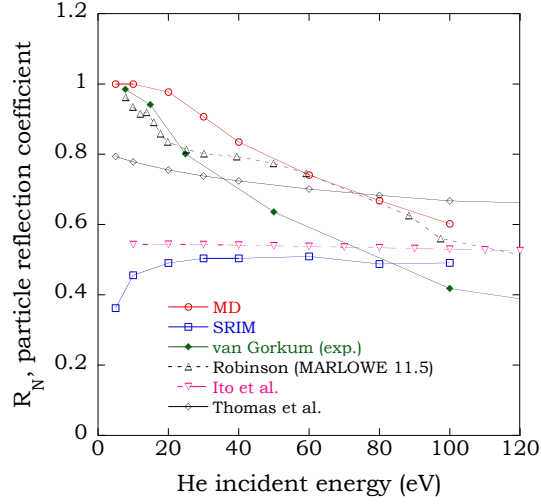}
\caption{\label{Rn} {Comparison of our MD simulation results for the
    dependence of particle reflection coefficient on He incident
    energy with some of the existing experimental and theoretical
    results. The MD simulation results are for normal deposition
    on a W(100) surface at $T = 1000K$. The SRIM simulation results
    correspond to a normal incidence of He atoms on 10000~\AA~thick W
    substrate. The experimental results by van Gorkum {\it et
      al}. correspond to normal deposition of He ions on a W(100)
    surface. The simulation results (MARLOWE 11.5) by Robinson
    correspond to normal deposition of He on a W(100) surface. The
    empirical formulas by Ito {\it et al}.  and Thomas {\it et al}.
    were used to produce the corresponding results shown.}}
\end{figure}

In Fig.~\ref{Rn} we compare our MD simulation results for incident
energy dependence of the particle reflection coefficient with some of
the existing experimental and theoretical results. Our MD simulation 
results are seen to be in qualitative agreement
with the experimental results by van Gorkum {\it et al}. and the
simulation results reported by Robinson (MARLOWE 11.5). This is to be
contrasted with those
obtained using the empirical formulas by Ito {\it et al}. and Thomas
{\it et al}. (Refs.\hspace{1mm}\onlinecite{Tabata, Eckstein, Thomas}),
as well as the results we obtained using the SRIM simulation
package\hspace{1mm}(see Ref.\hspace{1mm}\onlinecite{SRIM}), which
all have significant
disagreement.

\begin{figure}[t  ]
\includegraphics[width=7.0cm] {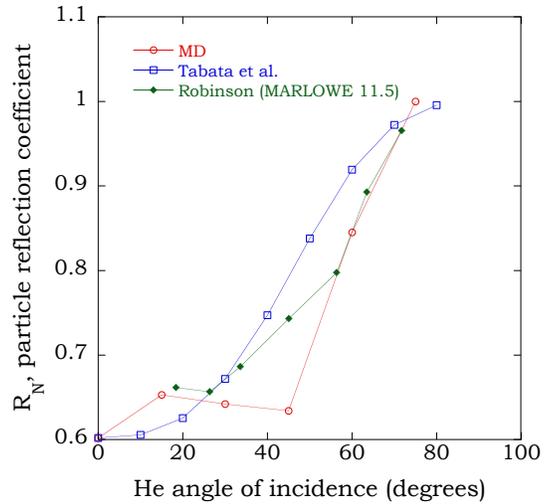}
\caption{\label{Rn-angle} {Comparison of our MD simulation results for
    the dependence of particle reflection coefficient on He incidence
    angle with some of the existing theoretical results. The MD
    simulation results correspond to deposition of He atoms on W(100)
    surface at $T = 1000K$. The incidence energy is 100~eV. The
    empirical formula by Tabata {\it et al}. is used to produce the
    corresponding results shown. The simulation results (MARLOWE 11.5)
    by Robinson correspond to deposition of He on a W(100) surface.}}
\end{figure}

Fig.~\ref{Rn-angle} compares our MD simulation results for the
incident angle dependence of the He particle reflection coefficient
with some of the existing theoretical results. Our MD simulation
results are in qualitative agreement with the results obtained using
the empirical formula by Tabata {\it et
  al}. (Ref.\hspace{1mm}\onlinecite{Tabata2}) and the simulation
results reported by Robinson
(Ref.\hspace{1mm}\onlinecite{Robinson}). We note that the enhanced
channeling of the He atoms incident at angles around $45^{\circ}$ 
(in the case of He deposition on the W(100) surface considered here)
leads to a significant deviation of our MD simulation results from the
those obtained using the empirical formula by Tabata {\it et
  al}. (Ref.\hspace{1mm}\onlinecite{Tabata2}) and the simulation
results reported by Robinson (Ref.\hspace{1mm}\onlinecite{Robinson}).

\begin{figure}[t  ]
\includegraphics[width=7.0cm] {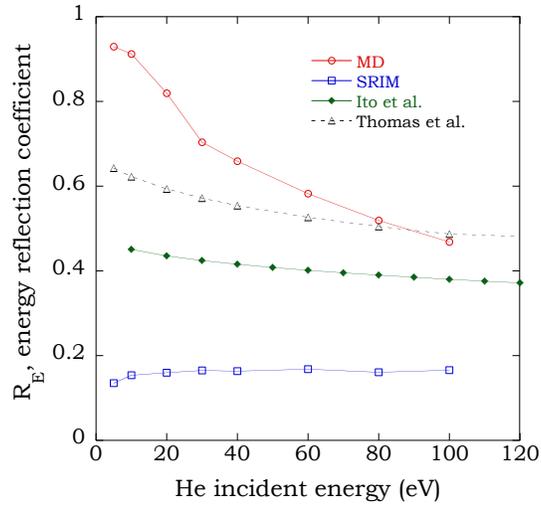}
\caption{\label{Re} {Comparison of our MD simulation results for a
    dependence of energy reflection coefficient on He incident energy
    with some of the existing experimental and theoretical
    results. The MD simulation results correspond to normal deposition
    on a W(100) surface at $T = 1000K$. The SRIM simulation results
    correspond to normal incidence of He atoms on 10000~\AA~thick W
    substrate. The empirical formulas by Ito {\it et al}.  and Thomas
    {\it et al}. are used to produce the corresponding results
    shown.}}
\end{figure}

Fig.~\ref{Re} compares our MD simulation results for the incident
energy dependence of the energy reflection coefficient with some of
the existing theoretical results. As can be seen from Fig.~\ref{Re},
our MD simulation results disagree with the results obtained
using the empirical formulas from Refs.~\onlinecite{Tabata},~\onlinecite{Thomas} (see also
Ref.~\onlinecite{Eckstein}). Our results are also in disagreement with
those obtained using the SRIM simulation package.

It should be noted that the empirical formulas by Ito {\it et
  al}. (see Ref.~\onlinecite{Tabata} and also
Ref.~\onlinecite{Eckstein}) and Thomas {\it et al}. (see
Ref.~\onlinecite{Thomas}) were fit to the existing experimental and
simulation (SRIM, MARLOWE) results corresponding to a very broad range
of He incident energies (up to 1000~KeV). This may help explain in
many cases, the disagreement of our MD simulation results with that of
the empirical formulas.

\begin{figure}[t  ]
\includegraphics[width=7.0cm] {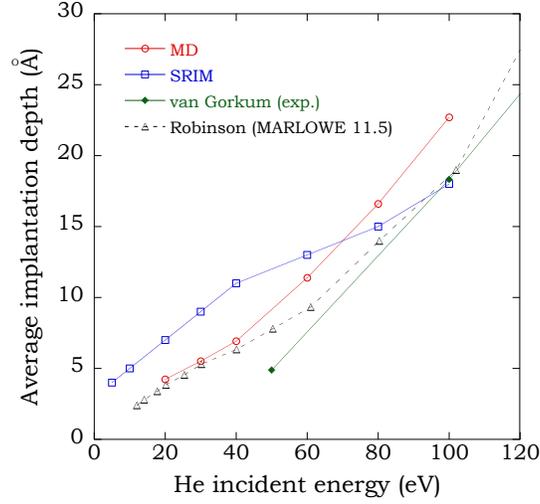}
\caption{\label{ID} {Comparison of our MD simulation results for the
    dependence of average implantation depth of He atoms on
    incidence He energy with some of the existing experimental and
    theoretical results. The MD simulation results correspond to
    normal deposition on a W(100) surface at $T = 1000K$. The SRIM
    simulation results correspond to normal incidence of He atoms on
    a 10000~\AA~thick W substrate. The experimental results by van
    Gorkum {\it et al}. correspond to normal deposition of He ions on
    a W(100) surface. The simulation results (MARLOWE 11.5) by Robinson
    correspond to normal deposition of He on a W(100) surface.}}
\end{figure}

In Fig.~\ref{ID} our MD simulation results for mean He implantation
depth are compared with some of the existing experimental and
theoretical results. The results from our simulations using the SRIM
simulation package are also shown. The MD simulation results are in
qualitative agreement with the experimental and theoretical results
shown in Fig.~\ref{ID} (Refs.~\onlinecite{vanGorkum, Robinson}). We
note that He stopping power in W is underestimated using classical MD
simulations (only the nuclear stopping is captured, but not the
electronic one). While the nuclear stopping dominates at lowest
incident energies of He atoms, at higher He incident energies
contribution from the electronic stopping becomes
appreciable~\cite{Haussalo}.

Surprisingly, some of our results disagree strongly with the MD
simulations results reported by Li {\it et al}.~(see
Ref.~\onlinecite{Li}), who used Derlet's potential for W
(Ref.~\onlinecite{Derlet}) in their studies. In addition, Li {\it et
  al}. developed and used in their studies a W-He pairwise
potential~\cite{Wang} different from the W-He pairwise potential 
recently developed by Juslin {\it et al}.~\cite{Juslin} (which we used
in our MD simulations). The reflection energy distributions we
obtained are very different than the results reported by Li
{\it et al}. In particular, we do not observe a significant dependence
of reflection energy distribution for reflected He atoms on the
temperature of the W(100) substrate (see Fig.~\ref{red-temp}). In
contrast, Li {\it et al}. reported a very strong temperature
dependence (see Fig. 5 of Ref.~\onlinecite{Li}). The implantation
depth distributions we obtained for different W(100) substrate
temperatures are also different, compared to the corresponding results
reported by Li {\it et al}. In particular, the distributions we report
have much longer tails (see Fig.~\ref{idd-temp}), compared to the
results shown in Fig. 6 of Ref.~\onlinecite{Li}.  It is unclear, other than the 
difference in interatomic potentials, what might have
contributed to these drastically different predictions.

\section{Conclusions\label{sec:conclusion}}

Recycling of low energy helium ions on tungsten surface occurs primarily
through reflection in a pure He plasma. This remains the case at
elevated temperature (1000 - 1500~K) for the tungsten substrate, as to
be expected in a fusion reactor like ITER, which is expected to have a pure
He plasma operational phase.  In contrast to the SRIM prediction but more
consistent with limited experimental data, MD
simulations reveal much larger particle and energy reflection
coefficients at low incidence ion energy (1-100~eV), which is the range for
ITER boundary plasmas by design.  This suggests that substantial He ion energy
will be fed back to the plasma instead of being absorbed by the
tungsten wall. The effect of that large energy feedback on boundary
plasmas will be pursued using the quantitative reflection results
reported here.

Of qualitative in addition to quantitative interest is the energy
distribution of the reflected He atoms, which undergoes an apparent
transition from an exponential distribution, Eq.~(\ref{eq:reflect-exp}),
to a Gaussian distribution, Eq.~(\ref{eq:reflect-gaussian}), as the
incident He energy is lowered from $E_i=80$~eV to $E_i=40$~eV. 
This poses a theoretical puzzle for future modeling exercises.
 The
angular distribution is found to have a weak azimuthal dependence, but a strong polar
dependence.  Although the conventional sine distribution is a
reasonable approximation for normal incidence, the $\theta_r$
dependence of reflected He has a more extreme form at large incidence angle $\theta_i.$

The range distribution of implanted He is of interest in setting up
longer-time scale materials studies under He irradiation.  To that
end, we recover the conventionally known Gaussian projected range
distribution with exponential tails.  The channeling effect is strong
at $T=300$~K, but becomes less pronounced at fusion-relevant
temperatures ($T=1000-1500$~K), although it still exists in producing
the exponential tail.  Again, our studies cover a range of incidence
angles, energies, W surface types, and substrate temperatures.

It is important to note that a detailed comparison of our new
results with existing experimental, empirical and simulation results
indicates much progress is desired for this decades-old problem.  The
agreements notwithstanding, the large spread in a key
quantity of interest, the particle reflection coefficient, between our
results and those of prior empirical and simulation data, is suggestive of the importance of 
additional experimental work in this important area of plasma-surface interaction research.

\acknowledgements

We wish to thank N.~Juslin for sharing the W and W-He interatomic potentials and 
T.~Tabata for alerting us about typos in the reproduction of their fitting formulas in Ref.~\cite{Eckstein}.
This work at Los Alamos National Laboratory (LANL)
was supported by the United States Department of Energy
(U.S. DOE), through the Office of Fusion Energy
Science (VB and XZT), and the Office of Basic Energy Sciences
(AFV). 
LANL is operated by Los Alamos National Security,
LLC, for the National Nuclear Security Administration
of the U.S. DOE, under contract DE-AC52-O6NA25396.

\begin{table*}
\tiny
  \caption{\label{table1} 
    Deposition of He ions on (100)W surface for 
    W substrate temperature $T=1000$~K. Incidence angles are: $0^{\circ}$, $15^{\circ}$, 
    $30^{\circ}$, $45^{\circ}$, $60^{\circ}$ and $75^{\circ}$ with respect to the substrate normal. 
    The first column shows the incident energy of the He atoms, the second column shows the 
    percentage of reflected He atoms, the third column shows the percentage of implanted He atoms, 
    the fourth column shows the total reflected energy (\% of total deposited energy), and the fifth column shows 
    the average implantation depth.} 

\begin{center}

\renewcommand{\arraystretch}{0.8}

\begin{tabular}{ccccc}
\hline
\multicolumn{5}{c}{Deposition angle is $0^{\circ}$} \\ \hline                       
$E_{ini}$~(eV) & \% refl. & \% impl. & E reflection coeff. (\%) & av. impl. depth~(\AA) \\ \hline 
5~(eV) & 100.0 & 0.0 & 92.93 & n/a \\ 
10~(eV) & 100.0 & 0.0 & 91.20 & n/a \\ 
20~(eV) & 97.7 & 2.3 & 81.93 & 4.23 \\ 
30~(eV) & 90.7 & 9.3 & 70.38 & 5.51 \\ 
40~(eV) & 83.5 & 16.5 & 65.89 & 6.93 \\ 
60~(eV) & 74.1 & 25.9 & 58.25 & 11.40 \\ 
80~(eV) & 66.8 & 33.2 & 51.91 & 16.60 \\ 
100~(eV) & 60.2 & 39.8 & 46.86 & 22.70 \\ \hline
\multicolumn{5}{c}{Deposition angle is $15^{\circ}$} \\ \hline
$E_{ini}$~(eV) & \% refl. & \% impl. & E reflection coeff. (\%) & av. impl. depth~(\AA) \\ \hline 
5~(eV) & 100.0 & 0.0 & 93.32 & n/a \\ 
10~(eV) & 100.0 & 0.0 & 91.11 & n/a \\ 
20~(eV) & 97.8 & 2.2 & 79.00 & 5.21 \\ 
30~(eV) & 88.9 & 11.1 & 66.08 & 5.0 \\ 
40~(eV) & 82.0 & 18.0 & 61.47 & 7.36 \\ 
60~(eV) & 72.0 & 28.0 & 55.43 & 10.57 \\ 
80~(eV) & 68.4 & 31.6 & 54.16 & 14.98 \\ 
100~(eV) & 65.3 & 34.7 & 51.65 & 20.09 \\ \hline
\multicolumn{5}{c}{Deposition angle is $30^{\circ}$} \\ \hline
$E_{ini}$~(eV) & \% refl. & \% impl. & E reflection coeff. (\%) & av. impl. depth~(\AA) \\ \hline 
5~(eV) & 100.0 & 0.0 & 93.66 & n/a \\ 
10~(eV) & 100.0 & 0.0 & 91.03 & n/a \\ 
20~(eV) & 98.3 & 1.7 & 78.47 & 3.97 \\ 
30~(eV) & 87.7 & 12.3 & 64.33 & 5.44 \\ 
40~(eV) & 79.5 & 20.5 & 60.61 & 6.47 \\ 
60~(eV) & 69.9 & 30.1 & 54.67 & 10.74 \\ 
80~(eV) & 63.5 & 36.5 & 48.80 & 15.54 \\ 
100~(eV) & 64.2 & 35.8 & 50.48 & 20.23 \\ \hline
\multicolumn{5}{c}{Deposition angle is $45^{\circ}$} \\ \hline
$E_{ini}$~(eV) & \% refl. & \% impl. & E reflection coeff. (\%) & av. impl. depth~(\AA) \\ \hline 
5~(eV) & 100.0 & 0.0 & 95.65 & n/a \\ 
10~(eV) & 100.0 & 0.0 & 94.35 & n/a \\ 
20~(eV) & 99.6 & 0.4 & 88.90 & 2.68 \\ 
30~(eV) & 94.1 & 5.9 & 79.20 & 4.41 \\ 
40~(eV) & 88.0 & 12.0 & 72.25 & 6.68 \\ 
60~(eV) & 73.6 & 26.4 & 60.57 & 10.23 \\ 
80~(eV) & 69.5 & 30.5 & 55.84 & 14.80 \\ 
100~(eV) & 63.4 & 36.6 & 50.07 & 20.29 \\ \hline
\multicolumn{5}{c}{Deposition angle is $60^{\circ}$} \\ \hline
$E_{ini}$~(eV) & \% refl. & \% impl. & E reflection coeff. (\%) & av. impl. depth~(\AA) \\ \hline 
5~(eV) & 100.0 & 0.0 & 98.29 & n/a \\ 
10~(eV) & 100.0 & 0.0 & 97.73 & n/a \\ 
20~(eV) & 100.0 & 0.0 & 97.28 & n/a \\ 
30~(eV) & 100.0 & 0.0 & 96.73 & n/a \\ 
40~(eV) & 99.5 & 0.5 & 94.87 & 4.59 \\ 
60~(eV) & 96.8 & 3.2 & 90.26 & 11.01 \\ 
80~(eV) & 92.7 & 7.3 & 84.74 & 14.36 \\ 
100~(eV) & 84.5 & 15.5 & 75.75 & 19.27 \\ \hline
\multicolumn{5}{c}{Deposition angle is $75^{\circ}$} \\ \hline
$E_{ini}$~(eV) & \% refl. & \% impl. & E reflection coeff. (\%) & av. impl. depth~(\AA) \\ \hline 
5~(eV) & 100.0 & 0.0 & 99.78 & n/a \\ 
10~(eV) & 100.0 & 0.0 & 99.74 & n/a \\ 
20~(eV) & 100.0 & 0.0 & 99.59 & n/a \\ 
30~(eV) & 100.0 & 0.0 & 99.54 & n/a \\ 
40~(eV) & 100.0 & 0.0 & 99.52 & n/a \\ 
60~(eV) & 100.0 & 0.0 & 99.46 & n/a \\ 
80~(eV) & 100.0 & 0.0 & 99.40 & n/a \\ 
100~(eV) & 100.0 & 0.0 & 99.33 & n/a \\ \hline

\end{tabular}
\end{center}
\end{table*}

\begin{table*}
  \caption{\label{table2} Deposition of He on (100)W surface for W substrate temperature $T=300$~K. 
} 

\begin{center}

\renewcommand{\arraystretch}{0.95}

\begin{tabular}{ccccc}
\hline
\multicolumn{5}{c}{Deposition angle is $0^{\circ}$} \\ \hline                         
$E_{ini}$~(eV) & \% refl. & \% impl. & E reflection coeff. (\%) & av. impl. depth~(\AA) \\ \hline 
5~(eV) & 100.0 & 0.0 & 93.02 & n/a \\ 
10~(eV) & 100.0 & 0.0 & 91.59 & n/a \\ 
20~(eV) & 97.0 & 3.0 & 82.39 & 4.47 \\ 
30~(eV) & 89.7 & 10.3 & 71.86 & 4.45 \\ 
40~(eV) & 86.7 & 13.3 & 69.05 & 6.94 \\ 
60~(eV) & 74.9 & 25.1 & 59.58 & 12.21 \\ 
80~(eV) & 66.3 & 33.7 & 51.90 & 22.18 \\ 
100~(eV) & 59.5 & 40.5 & 46.33 & 34.00 \\ \hline
\multicolumn{5}{c}{Deposition angle is $30^{\circ}$} \\ \hline
$E_{ini}$~(eV) & \% refl. & \% impl. & E reflection coeff. (\%) & av. impl. depth~(\AA) \\ \hline 
5~(eV) & 100.0 & 0.0 & 93.52 & n/a \\ 
10~(eV) & 100.0 & 0.0 & 91.38 & n/a \\ 
20~(eV) & 98.4 & 1.6 & 78.43 & 2.08 \\ 
30~(eV) & 87.1 & 12.9 & 64.61 & 4.76 \\ 
40~(eV) & 78.4 & 21.6 & 57.77 & 6.09 \\ 
60~(eV) & 68.9 & 31.1 & 53.67 & 10.65 \\ 
80~(eV) & 62.8 & 37.2 & 49.54 & 14.94 \\ 
100~(eV) & 64.2 & 35.8 & 50.09 & 21.12 \\ \hline
\multicolumn{5}{c}{Deposition angle is $60^{\circ}$} \\ \hline
$E_{ini}$~(eV) & \% refl. & \% impl. & E reflection coeff. (\%) & av. impl. depth~(\AA) \\ \hline 
5~(eV) & 100.0 & 0.0 & 98.20 & n/a \\ 
10~(eV) & 100.0 & 0.0 & 97.79 & n/a \\ 
20~(eV) & 100.0 & 0.0 & 97.42 & n/a \\ 
30~(eV) & 100.0 & 0.0 & 96.97 & n/a \\ 
40~(eV) & 99.9 & 0.1 & 96.13 & 6.92 \\ 
60~(eV) & 95.7 & 4.3 & 89.67 & 9.25 \\ 
80~(eV) & 92.8 & 7.2 & 85.13 & 12.67 \\ 
100~(eV) & 86.8 & 13.2 & 78.37 & 17.66 \\ \hline

\end{tabular}
\end{center}
\end{table*}

\begin{table*}
  \caption{\label{table3} Deposition of He on (100)W surface for W substrate temperature $T=1500$~K. 
} 

\begin{center}

\renewcommand{\arraystretch}{0.95}

\begin{tabular}{ccccc}
\hline
\multicolumn{5}{c}{Deposition angle is $0^{\circ}$} \\ \hline                         
$E_{ini}$~(eV) & \% refl. & \% impl. & E reflection coeff. (\%) & av. impl. depth~(\AA) \\ \hline
5~(eV) & 100.0 & 0.0 & 93.09 & n/a \\ 
10~(eV) & 100.0 & 0.0 & 91.09 & n/a \\ 
20~(eV) & 98.2 & 1.8 & 81.34 & 4.39 \\ 
30~(eV) & 89.9 & 10.1 & 69.28 & 5.20 \\ 
40~(eV) & 82.2 & 17.8 & 63.54 & 7.87 \\ 
60~(eV) & 74.5 & 25.5 & 58.39 & 11.00 \\ 
80~(eV) & 68.4 & 31.6 & 52.74 & 16.38 \\ 
100~(eV) & 64.0 & 36.0 & 50.13 & 22.33 \\ \hline
\multicolumn{5}{c}{Deposition angle is $30^{\circ}$} \\ \hline
$E_{ini}$~(eV) & \% refl. & \% impl. & E reflection coeff. (\%) & av. impl. depth~(\AA) \\ \hline 
5~(eV) & 100.0 & 0.0 & 94.25 & n/a \\ 
10~(eV) & 100.0 & 0.0 & 91.26 & n/a \\ 
20~(eV) & 98.3 & 1.7 & 77.69 & 3.10 \\ 
30~(eV) & 89.7 & 10.3 & 66.32 & 5.47 \\ 
40~(eV) & 77.9 & 22.1 & 59.00 & 7.27 \\ 
60~(eV) & 71.0 & 29.0 & 54.15 & 11.30 \\ 
80~(eV) & 66.8 & 33.2 & 52.34 & 16.13 \\ 
100~(eV) & 65.7 & 34.3 & 50.54 & 20.49 \\ \hline
\multicolumn{5}{c}{Deposition angle is $60^{\circ}$} \\ \hline
$E_{ini}$~(eV) & \% refl. & \% impl. & E reflection coeff. (\%) & av. impl. depth~(\AA) \\ \hline 
5~(eV) & 100.0 & 0.0 & 98.23 & n/a \\ 
10~(eV) & 100.0 & 0.0 & 97.85 & n/a \\ 
20~(eV) & 100.0 & 0.0 & 97.32 & n/a \\ 
30~(eV) & 99.7 & 0.3 & 96.39 & 6.49 \\ 
40~(eV) & 99.3 & 0.7 & 94.58 & 7.21 \\ 
60~(eV) & 94.6 & 5.4 & 87.58 & 9.81 \\ 
80~(eV) & 88.8 & 11.2 & 81.22 & 15.09 \\ 
100~(eV) & 86.2 & 13.8 & 76.47 & 19.35 \\ \hline

\end{tabular}
\end{center}
\end{table*}

\begin{table*}
\caption{\label{table4} Deposition of He on (110)W surface for W substrate temperature $T=1000$~K. }
\begin{center}

\renewcommand{\arraystretch}{0.95}

\begin{tabular}{ccccc}
\hline
\multicolumn{5}{c}{Deposition angle is $0^{\circ}$} \\ \hline                         
$E_{ini}$~(eV) & \% refl. & \% impl. & E reflection coeff. (\%) & av. impl. depth~(\AA) \\ \hline 
5~(eV) & 100.0 & 0.0 & 93.10 & n/a \\ 
10~(eV) & 100.0 & 0.0 & 89.68 & n/a \\ 
20~(eV) & 96.5 & 3.5 & 66.44 & 3.85 \\ 
30~(eV) & 85.5 & 14.5 & 59.42 & 4.87 \\ 
40~(eV) & 77.8 & 22.2 & 57.42 & 6.56 \\ 
60~(eV) & 72.2 & 27.8 & 54.77 & 10.01 \\ 
80~(eV) & 62.6 & 37.4 & 48.34 & 13.55 \\ 
100~(eV) & 61.4 & 38.6 & 46.19 & 19.13 \\ \hline
\multicolumn{5}{c}{Deposition angle is $30^{\circ}$} \\ \hline
$E_{ini}$~(eV) & \% refl. & \% impl. & E reflection coeff. (\%) & av. impl. depth~(\AA) \\ \hline 
5~(eV) & 100.0 & 0.0 & 94.74 & n/a \\ 
10~(eV) & 100.0 & 0.0 & 93.84 & n/a \\ 
20~(eV) & 99.2 & 0.8 & 84.62 & 3.73 \\ 
30~(eV) & 87.6 & 12.4 & 66.63 & 4.78 \\ 
40~(eV) & 76.7 & 23.3 & 58.05 & 6.46 \\ 
60~(eV) & 67.7 & 32.3 & 50.58 & 10.48 \\ 
80~(eV) & 60.7 & 39.3 & 46.45 & 14.90 \\ 
100~(eV) & 57.9 & 42.1 & 44.51 & 21.43 \\ \hline
\multicolumn{5}{c}{Deposition angle is $60^{\circ}$} \\ \hline
$E_{ini}$~(eV) & \% refl. & \% impl. & E reflection coeff. (\%) & av. impl. depth~(\AA) \\ \hline 
5~(eV) & 100.0 & 0.0 & 98.52 & n/a \\ 
10~(eV) & 100.0 & 0.0 & 98.27 & n/a \\ 
20~(eV) & 100.0 & 0.0 & 98.12 & n/a \\ 
30~(eV) & 100.0 & 0.0 & 97.92 & n/a \\ 
40~(eV) & 100.0 & 0.0 & 97.60 & n/a \\ 
60~(eV) & 99.7 & 0.3 & 96.51 & 2.05 \\ 
80~(eV) & 97.5 & 2.5 & 92.67 & 14.05 \\ 
100~(eV) & 94.0 & 6.0 & 87.36 & 14.95 \\ \hline

\end{tabular}
\end{center}
\end{table*}

\begin{table*}
  \caption{\label{table5} Deposition of He on (310)W surface for W substrate temperature $T=1000$~K. }
\begin{center}

\renewcommand{\arraystretch}{0.95}

\begin{tabular}{ccccc}
\hline
\multicolumn{5}{c}{Deposition angle is $0^{\circ}$} \\ \hline                         
$E_{ini}$~(eV) & \% refl. & \% impl. & E reflection coeff. (\%) & av. impl. depth~(\AA) \\ \hline 
5~(eV) & 100.0 & 0.0 & 93.30 & n/a \\ 
10~(eV) & 100.0 & 0.0 & 90.85 & n/a \\ 
20~(eV) & 98.2 & 1.8 & 75.26 & 4.08 \\ 
30~(eV) & 85.0 & 15.0 & 60.48 & 4.76 \\ 
40~(eV) & 76.4 & 23.6 & 55.63 & 6.90 \\ 
60~(eV) & 70.4 & 29.6 & 54.26 & 10.62 \\ 
80~(eV) & 67.8 & 32.2 & 52.89 & 15.49 \\ 
100~(eV) & 63.3 & 36.7 & 49.18 & 20.80 \\ \hline
\multicolumn{5}{c}{Deposition angle is $30^{\circ}$} \\ \hline
$E_{ini}$~(eV) & \% refl. & \% impl. & E reflection coeff. (\%) & av. impl. depth~(\AA) \\ \hline 
5~(eV) & 100.0 & 0.0 & 94.54 & n/a \\ 
10~(eV) & 100.0 & 0.0 & 92.93 & n/a \\ 
20~(eV) & 99.1 & 0.9 & 82.82 & 3.71 \\ 
30~(eV) & 91.7 & 8.3 & 71.06 & 5.31 \\ 
40~(eV) & 81.9 & 18.1 & 65.36 & 6.57 \\ 
60~(eV) & 73.3 & 26.7 & 58.49 & 11.44 \\ 
80~(eV) & 69.0 & 31.0 & 55.09 & 16.69 \\ 
100~(eV) & 65.0 & 35.0 & 50.56 & 21.47 \\ \hline
\multicolumn{5}{c}{Deposition angle is $60^{\circ}$} \\ \hline
$E_{ini}$~(eV) & \% refl. & \% impl. & E reflection coeff. (\%) & av. impl. depth~(\AA) \\ \hline 
5~(eV) & 100.0 & 0.0 & 97.74 & n/a \\ 
10~(eV) & 100.0 & 0.0 & 97.18 & n/a \\ 
20~(eV) & 100.0 & 0.0 & 96.60 & n/a \\ 
30~(eV) & 99.6 & 0.4 & 94.57 & 3.46 \\ 
40~(eV) & 96.9 & 3.1 & 90.99 & 6.53 \\ 
60~(eV) & 91.4 & 8.6 & 83.67 & 11.72 \\ 
80~(eV) & 85.2 & 14.8 & 76.29 & 15.92 \\ 
100~(eV) & 83.6 & 16.4 & 73.43 & 21.02 \\ \hline

\end{tabular}
\end{center}
\end{table*}

\clearpage

\begin{table*}
  \caption{\label{table6} Deposition of He on W surface studied using the 
    SRIM simulation package~\cite{SRIM}. Incidence angle is $0^{\circ}$ 
    with respect to substrate normal. Depth of the W target is 10000\AA. 
    The results correspond to an average over 10000 depositions. 
    The first column shows the incident energy of the He atoms, the second column shows the 
    percentage of reflected He atoms, the third column shows the percentage of implanted He atoms, 
    the fourth column shows the total reflected energy (\% of total deposited energy), and the fifth column shows 
    the average implantation depth.} 

\begin{center}
\begin{tabular}{ccccc}
\hline
\multicolumn{5}{c}{Deposition angle is $0^{\circ}$} \\ \hline                         
$E_{ini}$~(eV) & \% refl. & \% impl. & E reflection coeff. (\%) & av. impl. depth~(\AA) \\ \hline
5~(eV) & 36.22 & 63.78 & 13.49 & 4.0 \\ 
10~(eV) & 45.60 & 54.40 & 15.39 & 5.0 \\ 
20~(eV) & 49.05 & 50.95 & 15.96 & 7.0 \\ 
30~(eV) & 50.37 & 49.63 & 16.54 & 9.0 \\ 
40~(eV) & 50.37 & 49.63 & 16.34 & 11.0 \\ 
60~(eV) & 50.95 & 49.05 & 16.81 & 13.0 \\ 
80~(eV) & 48.76 & 51.24 & 16.07 & 15.0 \\ 
100~(eV) & 49.12 & 50.88 & 16.58 & 18.0 \\ \hline

\end{tabular}
\end{center}
\end{table*}

\end{document}